\documentclass[aps,prd,twocolumn,superscriptaddress,preprintnumbers,showkeys,nofootinbib,longbibliography]{revtex4-2}  
\usepackage[latin1]{inputenc}                  
\usepackage{graphicx}  
\usepackage{bm}       
\usepackage{textcomp}
\usepackage{epsfig}
\usepackage{epsf}
\usepackage{bbm}

\usepackage{latexsym}                        
\usepackage{amsfonts}                           
\usepackage{amssymb}                            
\usepackage{amsmath}                            
\usepackage[mathscr]{eucal}                    
\usepackage{dcolumn}                             
\usepackage{multirow}                             
\usepackage{bm}                                 
\usepackage{hyperref}                         
\usepackage[usenames]{color}
\usepackage{array}
\usepackage{appendix}
\usepackage{bbold}
\usepackage{soul,xcolor}
\usepackage{lipsum}
\usepackage{braket}
\setstcolor{red}
\usepackage{float}
\usepackage{booktabs}
\usepackage{slashed}
\usepackage[normalem]{ulem}

\newcommand{\be}{\begin{equation}}
\newcommand{\ee}{\end{equation}}
\newcommand{\bea}{\begin{eqnarray}}
\newcommand{\eea}{\end{eqnarray}}

\def \3{\ss }

\newcommand{\tr}{\operatorname{Tr}}

\newcommand{\beq}{\begin{eqnarray}}
\newcommand{\eeq}{\end{eqnarray}}

\renewcommand{\arraystretch}{1.8}

\newcommand{\ord}{\mathcal{O}}

\newcommand{\gfi}{\gamma_5}

\newcommand{\jint}{\mathcal{J}}
\newcommand{\jintb}{\bar{\mathcal{J}}}

\newcommand{\gaumd}{g_A^{u-d}}
\newcommand{\gaupd}{g_A^{u+d}}

\newcommand{\tins}{t_\text{ins}}

\usepackage{letltxmacro}

\newif\ifcorrectingmode
\correctingmodetrue

\begin{document}

\title{Flavor decomposition of the nucleon axial and tensor charges from lattice QCD }

\author{C.~Alexandrou} \affiliation{Department of Physics, University of Cyprus, P.O. Box 20537, 1678 Nicosia, Cyprus} \affiliation{Computation-based Science and Technology Research Center, The Cyprus Institute, 20 Kavafi Str., Nicosia 2121, Cyprus}
\author{S.~Bacchio} \affiliation{Computation-based Science and Technology Research Center, The Cyprus Institute, 20 Kavafi Str., Nicosia 2121, Cyprus}
 \author{M.~Di Carlo} \affiliation{Istituto Nazionale di Fisica Nucleare, Sezione di Roma Tor Vergata, Via della Ricerca Scientifica 1, I-00133 Roma, Italy}
\author{A.~Evangelista} \affiliation{Department of Physics, University of Cyprus, P.O. Box 20537, 1678 Nicosia, Cyprus}
\author{C.~Iona} \affiliation{Department of Physics, University of Cyprus, P.O. Box 20537, 1678 Nicosia, Cyprus}\affiliation{Computation-based Science and Technology Research Center, The Cyprus Institute, 20 Kavafi Str., Nicosia 2121, Cyprus}
\author{F.~Pittler} \affiliation{Computation-based Science and Technology Research Center, The Cyprus Institute, 20 Kavafi Str., Nicosia 2121, Cyprus}
\author{G.~Spanoudes} \affiliation{Department of Physics, University of Cyprus, P.O. Box 20537, 1678 Nicosia, Cyprus}

\begin{abstract}
\vspace{2pt}
\centerline{\includegraphics[width=0.13\textwidth]{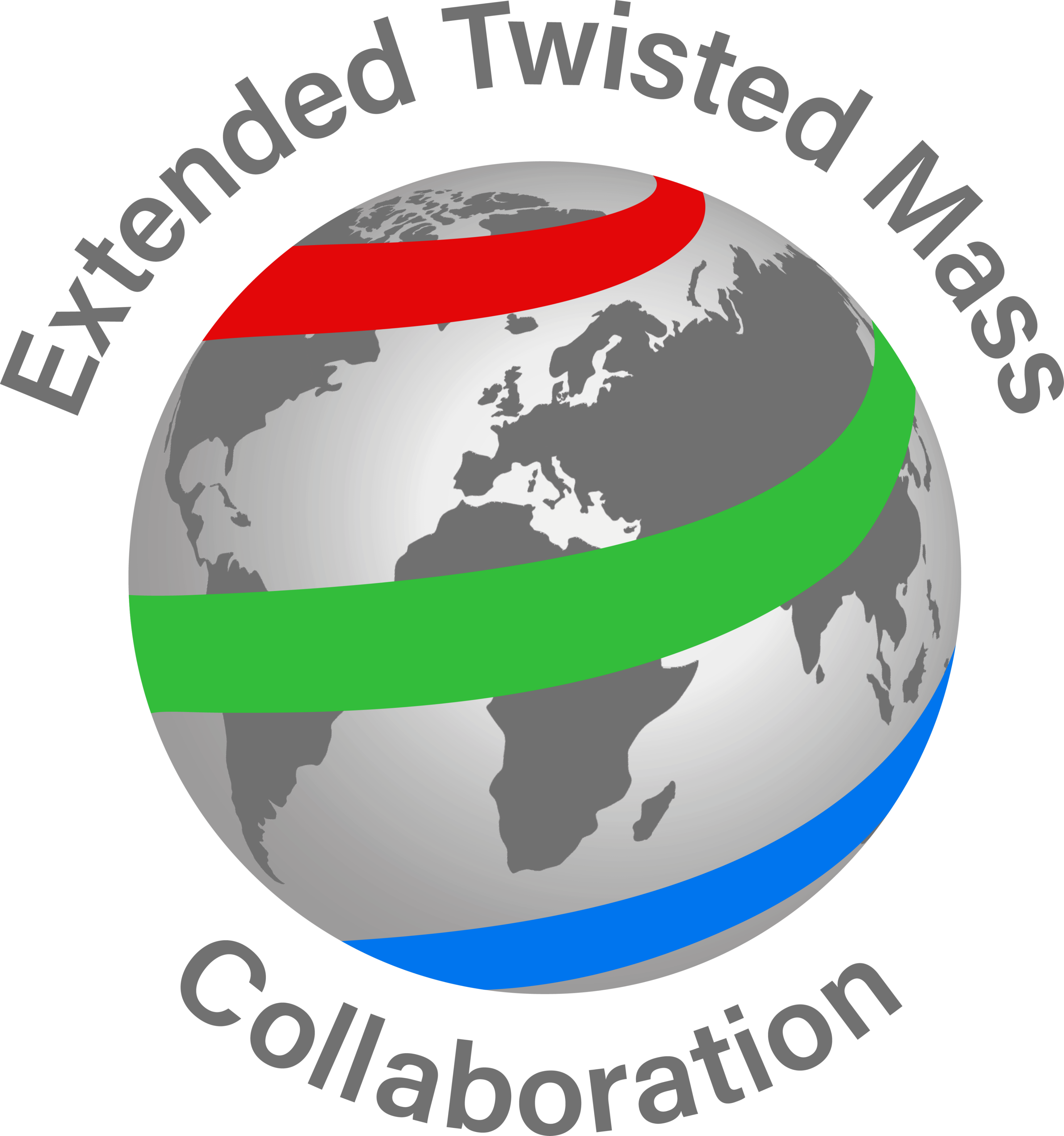}} 
\vspace{10pt}
We determine the nucleon axial and tensor charges using four ensembles of twisted-mass clover-improved fermions with approximately equal physical volumes of about 5.5~fm and four lattice spacings, $ a=0.05$~fm, 0.06~fm, 0.07~fm, and 0.08~fm. The mass of the degenerate up and down, strange and charm quarks are tuned to approximately their physical values. We compute both connected and disconnected contributions to determine the isovector and isoscalar charges and their flavour decomposition. For the nucleon isovector axial charge we find $g_A^{u-d}=1.263(34)(26)$, in agreement with the experimental value. For the isoscalar axial charge  we find $g_A^{u+d}=0.458(16)(1)$, while for the corresponding tensor charges we find $g_T^{u-d}=0.951(21)(5)$ and $g_T^{u+d}=0.556(16)(5)$.
\end{abstract}


\maketitle

\section{Introduction}
The nucleon axial and tensor charges are fundamental quantities in the Standard Model and provide important insights into nucleon structure and dynamics. The axial charge, $\gaumd$, determines the strength of neutron beta decay and provides a direct probe of chiral symmetry breaking in hadron physics. It also enters the interpretation of neutrinoless double-beta decay and plays a central role in precision tests of the unitarity of the Cabibbo--Kobayashi--Maskawa (CKM) matrix. Flavor-diagonal axial charges, $g_A^f$, characterize the contribution of quark flavor $f$ to the intrinsic spin of the nucleon, $\frac{1}{2}\Delta\Sigma_f$, and are constrained by deep-inelastic scattering (DIS) experiments at facilities, such as Jefferson Lab and CERN. Extending these measurements to the low-Bjorken-$x$ region is one of the key physics goals of the future Electron-Ion Collider (EIC), where improved sensitivity to the quark and gluon spin structure of the nucleon is expected.

The isovector tensor charge, $g_T^{u-d}$, is less well known but plays an important role in constraining beyond-the-Standard-Model (BSM) interactions~\cite{Bhattacharya:2011qm}. It provides essential theoretical input for the interpretation of results from neutrino scattering experiments, including DUNE~\cite{Bischer:2018zcz}, COHERENT~\cite{COHERENT:2017ipa}, and GEMMA~\cite{Beda:2012zz}, as well as from direct dark matter detection experiments~\cite{Undagoitia:2015gya,Chizhov:1995wp,Dhargyal:2016jgo}. An accurate determination of the tensor charge is also important for phenomenological studies of the nucleon transversity parton distribution function~\cite{Cocuzza:2023oam}, where it can be used as input in such analyses.

In this work, we compute the nucleon axial and tensor charges using twisted-mass, clover-improved fermion ensembles at four different lattice spacings. The ensembles are generated by the Extended Twisted Mass Collaboration (ETMC) with a mass degenerate up and down quark doublet and strange and charm quarks ($N_f=2+1+1$), with all quark masses tuned to approximately their physical values. This setup, which we refer to as the physical-mass point, allows for a precise determination of both isovector and flavor-diagonal charges without the need for chiral extrapolations. Our results provide important benchmarks for lattice QCD calculations and deliver essential theoretical input for precision studies probing BSM interactions, CP violation, and dark-matter interactions.

\section{Methodology}
\label{sec:methodology}
The nucleon charges for each quark flavor $f$, denoted as $g^f_{A,T}$, are derived from the nucleon matrix elements of the corresponding axial and tensor operators at zero momentum transfer. They are defined as

\begin{equation}
\langle N|\bar{\psi}^f\Gamma_{A,T}\psi^f|N\rangle = g^f_{A,T} \bar{u}_N\Gamma_{A,T} u _N\,,
\end{equation}
where $u_N$ is the nucleon spinor and $\Gamma_A=\gamma_\mu\gamma_5$ for the axial-vector and $\Gamma_T=\sigma_{\mu\nu}$ for the tensor operators. 

\subsection{Gauge ensembles and statistics}
\label{sec:stats}

\begin{table}[h!]
	\centering
    \caption{Parameters of the $N_f=2+1+1 $ ensembles analyzed in this work. In the first column, we give the name of
             the ensemble, in the second the abbreviated name, in the third the lattice volume, in the fourth $\beta = 6/g^2$ with $g$ the bare coupling constant, in the fifth
             the lattice spacing and in the last the pion mass. Lattice spacings and pion masses are
             taken from Ref.~\cite{ExtendedTwistedMass:2024nyi}.}
	{\small
		\renewcommand{\arraystretch}{1.2}
		\renewcommand{\tabcolsep}{1.5pt}
        \begin{tabular}{c|c|c|c|c|c}
        \hline \hline
        Ensemble & Abrv. & $V/a^4$ & $\beta$ & $a$~[fm] & $m_\pi$~[MeV] \\ \hline
        cB211.072.64  & B64  & $64^3 \times 128$  & 1.778 & 0.07948(11) & 140.2(2) \\
        cC211.060.80  & C80  & $80^3 \times 160$  & 1.836 & 0.06819(14) & 136.7(2) \\
        cD211.054.96  & D96  & $96^3 \times 192$  & 1.900 & 0.056850(90)& 140.8(2) \\
        cE211.044.112 & E112 & $112^3 \times 224$ & 1.960 & 0.04892(11) & 136.5(2) \\
        \hline \hline
        \end{tabular}}
    \label{tbl:Ensembles}
\end{table}

To compute the nucleon matrix elements, we examine gauge configurations 
produced using the twisted-mass fermion discretization, which provides 
automatic \(\ord(a)\) improvement~\cite{Frezzotti:2000nk,Frezzotti:2003ni}. 
The action incorporates a clover term~\cite{Sheikholeslami:1985ij}, which helps suppress isospin-breaking effects associated with the twisted mass fermion formulation.

The isosymmetric pion mass, \(m_\pi = 135\)~MeV~\cite{Alexandrou:2018egz,Finkenrath:2022eon}, 
is reproduced by tuning the bare light quark parameter \(\mu_l\). For the heavy quarks, the parameters \(\mu_s\) and \(\mu_c\) are tuned using the kaon mass, a suitably defined ratio of the D-meson mass to its decay constant, and a phenomenologically motivated ratio between the strange and charm quark masses, following Refs.~\cite{Alexandrou:2018egz,Finkenrath:2022eon}. Table~\ref{tbl:Ensembles} lists the parameters of the ensembles employed in this study. The lattice spacings extracted from the meson sector and the pion masses are taken from Ref.~\cite{ExtendedTwistedMass:2024nyi}, with the former being consistent with those obtained from the nucleon sector in Ref.~\cite{ExtendedTwistedMass:2021gbo}.

\begin{widetext}

\begin{table*}[h!]
\caption{Statistics used for the connected two- and three-point
    functions.  In the first row we give the name of the ensemble and in the second the number of configurations analyzed per ensemble. For each ensemble,  we provide the sink-source time separations in lattice
    units (first column) and physical units (second column) and the
    number of source positions per configuration (third column). The last row indicates the number of source
    positions used for the two-point functions. }
\label{tbl:Stats}
\begin{ruledtabular}
\renewcommand{\arraystretch}{1.0} 
\begin{tabular}{c ccc ccc ccc ccc}
    & \multicolumn{3}{c}{{cB211.072.64}} 
    & \multicolumn{3}{c}{{cC211.060.80}} 
    & \multicolumn{3}{c}{{cD211.054.96}}
    & \multicolumn{3}{c}{{cE211.044.112}} \\ 
    & \multicolumn{3}{c}{749 configurations} 
    & \multicolumn{3}{c}{400 configurations} 
    & \multicolumn{3}{c}{496 configurations}
    & \multicolumn{3}{c}{478 configurations} \\ 
    \cmidrule(lr){2-4} \cmidrule(lr){5-7} \cmidrule(lr){8-10} \cmidrule(lr){11-13} 
    & $t_s/a$ & $t_s$ [fm] & $n_{\rm src}$  
    & $t_s/a$ & $t_s$ [fm] & $n_{\rm src}$  
    & $t_s/a$ & $t_s$ [fm] & $n_{\rm src}$
    & $t_s/a$ & $t_s$ [fm] & $n_{\rm src}$  \\ 
    \hline
    & 8  & 0.64 & 1   & 6  & 0.41 & 1   & 8  & 0.46 & 1  & 8  & 0.39 & 1   \\
    & 10 & 0.80 & 2   & 8  & 0.55 & 2   & 10 & 0.57 & 2  & 11 & 0.54 & 2   \\
    & 12 & 0.96 & 5   & 10 & 0.69 & 4   & 12 & 0.68 & 4  & 14 & 0.69 & 4   \\
    & 14 & 1.12 & 10  & 12 & 0.82 & 10  & 14 & 0.80 & 8  & 17 & 0.83 & 8   \\
    & 16 & 1.28 & 32  & 14 & 0.96 & 22  & 16 & 0.91 & 16 & 20 & 0.98 & 16  \\
    & 18 & 1.44 & 112 & 16 & 1.10 & 48  & 18 & 1.03 & 32 & 23 & 1.13 & 32  \\
    & 20 & 1.60 & 128 & 18 & 1.24 & 45  & 20 & 1.14 & 64 & 26 & 1.27 & 64  \\
    &    &      &     & 20 & 1.37 & 116 & 22 & 1.25 & 16 & 29 & 1.42 & 128 \\
    &    &      &     & 22 & 1.51 & 246 & 24 & 1.37 & 32 &    &      &     \\
    &    &      &     &    &      &     & 26 & 1.48 & 64 &    &      &     \\
    \hline
    & \multicolumn{3}{c}{Nucleon 2pt $n_{\rm src}$=349} 
    & \multicolumn{3}{c}{Nucleon 2pt $n_{\rm src}$=650} 
    & \multicolumn{3}{c}{Nucleon 2pt $n_{\rm src}$=368}
    & \multicolumn{3}{c}{Nucleon 2pt $n_{\rm src}$=467} \\
\end{tabular}
\end{ruledtabular}
\end{table*}
\vspace{-0.5cm}
\begin{table*}[h!]
\caption{Statistics used for the disconnected loops. In the first row we give the name of the ensemble and in the second the number of configurations analyzed per ensemble. For each
  ensemble, we give  from left to right: i) the number
  of deflated eigenvectors $N_{\rm defl}$, ii) the number of stochastic
  sources $N_r$, iii) the number of Hadamard vectors $N_{\rm Had}$, and
  iv) the total number of computed vectors, $N_{\rm vect}$, which after
  color and spin dilution are obtained via $12 \times N_ r \times
  N_{\rm Had}$.} \label{tbl:Stats-disc}
\begin{ruledtabular}
\renewcommand{\arraystretch}{1.0}
\begin{tabular}{c cccc cccc cccc cccc}
    & \multicolumn{4}{c}{{cB211.072.64}} 
    & \multicolumn{4}{c}{{cC211.060.80}} 
    & \multicolumn{4}{c}{{cD211.054.96}}
    & \multicolumn{4}{c}{{cE211.044.112}} \\ 
    & \multicolumn{4}{c}{749 configurations} 
    & \multicolumn{4}{c}{400 configurations} 
    & \multicolumn{4}{c}{494 configurations}
    & \multicolumn{4}{c}{450 configurations} \\ 
    \cmidrule(lr){2-5} \cmidrule(lr){6-9} \cmidrule(lr){10-13} \cmidrule(lr){14-17} 
    {Flavor} 
    & $N_{\text{defl}}$ & $N_r$ & $N_{\text{Had}}$ & $N_{\text{vect}}$  
    & $N_{\text{defl}}$ & $N_r$ & $N_{\text{Had}}$ & $N_{\text{vect}}$  
    & $N_{\text{defl}}$ & $N_r$ & $N_{\text{Had}}$ & $N_{\text{vect}}$
    & $N_{\text{defl}}$ & $N_r$ & $N_{\text{Had}}$ & $N_{\text{vect}}$  \\ 
    \hline
    Light
    & 200 & 1  & 512 & 6144
    & 450 & 1  & 512 & 6144
    & 0   & 8  & 512 & 49152
    & 530 & 1  & 512 & 6144  \\
    Strange
    & 0   & 2  & 512 & 12288
    & 0   & 4  & 512 & 24576
    & 0   & 4  & 512 & 24576
    & 0   & 2  & 512 & 12288 \\
    Charm
    & 0   & 12 & 32  & 4608
    & 0   & 1  & 512 & 6144
    & 0   & 1  & 512 & 6144
    & 0   & 1  & 512 & 6144  \\
\end{tabular}
\end{ruledtabular}

\end{table*}

\end{widetext}

In order to extract the nucleon charges, we compute two- and three-point 
nucleon correlation functions and perform fits to isolate the desired 
ground-state matrix element, as described in detail in 
Sec.~\ref{sec:FitProc}. In Tables~\ref{tbl:Stats} and \ref{tbl:Stats-disc}, we present the statistics used for the calculation of the correlation functions, providing  the number of configurations analyzed, the number of source positions per configuration as well as details on the stochastic noise and Hadamard vectors used in the computation of disconnected  quark loops. The connected three-point functions are evaluated using the so-called 
\textit{fixed sink} approach~\cite{Detmold:2001jb,Bernard:1985ss}, in which a separate inversion is performed for every source-sink separation $t_s$. As reported in Table~\ref{tbl:Stats}, the number of source positions per configuration is increased for larger values of  \(t_s\), ensuring that the statistical uncertainties of the connected three-point functions remain approximately constant.

The disconnected quark loops are calculated for the light, strange,
and charm quarks. To enhance the signal-to-noise ratio of these
disconnected loops, several noise-reduction techniques are
applied. These include the one-end trick~\cite{McNeile:2006bz}, exact
deflation of low modes~\cite{Gambhir:2016uwp,Alexandrou:2026rca}, and hierarchical
probing~\cite{Stathopoulos:2013aci}, as explained in Ref.~\cite{Abdel-Rehim:2016pjw} for the specific application to twisted mass fermions. The one-end trick is applied to all quark loops for all four ensembles.  We also use hierarchical probing for all quark loops with a probing distance of 8 for the light and strange quark loops and with 4 for the charm-quark
loops.  Deflation of low modes is applied to the light quark loops for
the cB211.072.64 and cC211.060.80  ensembles. For the cD211.054.96
ensemble,   a higher number of stochastic sources are used since using
deflation for this 96$^3\times$192 lattice, which scales with the
square of the volume, was prohibitively expensive  with our original implementation and the then accessible hardware. With our improved implementation and latest hardware we are able to employ deflation for the cE211.044.112 ensemble.

\subsection{Computation of correlators}
\label{sec:correlation functions}
The determination of the nucleon matrix elements requires the evaluation 
of both three- and two-point Euclidean time correlation functions. The 
two-point function is given by
\begin{equation}
    \begin{aligned}
        C(\Gamma_0,\vec{p};t_s) &= \sum_{\vec{x}_s}e^{-i\vec{x}_s\cdot \vec{p}} \times \\
                                    & \tr \left[ \Gamma_0 \braket{\jint_N(t_s,\vec{x}_s) \jintb_N (0,\vec{0}) } \right],
    \end{aligned}
\end{equation}
where we take the source at the origin, $t_s, \vec{x}_s$ are the sink
coordinates, $\Gamma_0$ is the unpolarized, positive parity projector
$\Gamma_0=\frac{1}{2}(1+\gamma_0)$ and the interpolating field for the
nucleon is
\begin{gather} \label{eq:IntField}
    \jint_N(t,\vec{x})=\epsilon_{abc}u_a(x) \left( u^T_b(x) C \gfi d_c(x) \right).
\end{gather}
To enhance the overlap of the interpolating field with the nucleon 
ground state and suppress excited-state contamination, thereby allowing 
the ground state to dominate at the smallest possible time separations, 
we apply \textit{Gaussian smearing}~\cite{Alexandrou:1992ti,Gusken:1989qx} 
to the quark fields that enter the interpolating field.
\begin{gather}
    \widetilde{\psi}(\vec{x},t)=\sum_{\vec{y}}[1+a_G H(\vec{x},\vec{y};U(t))]^{N_G} \psi(\vec{y},t),
\end{gather}
where $\psi$ is either $u$ or $d$ of Eq.~(\ref{eq:IntField}) and $H$ is the hopping matrix 
\begin{gather}
    H(\vec{x},\vec{y};U(t))=\sum_{i=1}^{3}\left[ U_i(x)\delta_{x,y-\hat{i}}+U_i^\dagger(x-\hat{i})\delta_{x,y+\hat{i}} \right].
\end{gather}
The parameters $\alpha_G$ and $N_G$ are adjusted~\cite{Alexandrou:2018sjm,Alexandrou:2019ali} to achieve a nucleon smearing radius of approximately 0.5~fm. To suppress statistical noise arising from ultraviolet fluctuations, APE smearing~\cite{Albanese:1987ds} is applied to the gauge links used in the hopping matrix.

The three-point function for e.g.  the axial-vector   for an arbitrary momentum transfer $\vec{q}$ is given by
\begin{gather}
        C_\mu(\Gamma_k,\vec{q},\vec{p^\prime};t_s,\tins)=\sum_{\vec{x}_s,\vec{x}_{\rm ins}} e^{i \vec{x}_{\rm ins} \cdot \vec{q}} e^{-i \vec{x}_s \cdot \vec{p^\prime}} \times \notag \\ 
       \tr \left[ \Gamma_k \braket{\jint_N(t_s,\vec{x}_s) {\cal A}_\mu(\tins,\vec{x}_{\rm ins}) \jintb_N (0,\vec{0}) } \right],
\end{gather}
where $\Gamma_k=i \Gamma_0 \gfi \gamma_k$ is the polarized projector, $\vec{p}^{\,\prime}$ is the sink momentum and $\vec{q}=-\vec{p}^{\,\prime}$,
${\cal A}_\mu$ is the axial operator, which is defined as 
\begin{gather}
    {\cal A}_\mu=\bar{\psi}\gamma_\mu\gamma_5\psi.
\end{gather}
Analogous expressions hold for the tensor three-point functions, where the tensor operator is 
\begin{gather}
    {\cal T}_{\mu\nu}=\bar{\psi}\sigma_{\mu\nu}\psi,
\end{gather}
with $\sigma_{\mu\nu}=\frac{i}{2}[\gamma_\mu,\gamma_\nu]$. These tensor matrix elements are obtained using the same polarized projector $\Gamma_k$. In the following, we will adopt the axial-vector case for our discussion, 
noting that the generalization to tensor operator insertions is 
straightforward. Since we extract nucleon charges in the forward limit, we take $\vec{q}=0$ and work in the rest frame of the nucleon.

\subsection{Analysis of correlators}
\label{sec:analysis}
The spectral decomposition of the two- and three-point functions 
is given respectively by 
\begin{gather}\label{eq:Twop}
    C(\Gamma_0,\vec{0};t_s)= \sum_{i=0}^{\infty} c_i e^{-E_i\,t_s}~~\text{and} \\
    C_\mu(\Gamma_k,\vec{0};t_s,\tins)=\!\!\!\!\!\sum_{i=0,j=0}^{\infty}\!\!\!\!\! A^{i,j}_\mu (\Gamma_k)e^{-E_i(t_s-\tins)-E_j\tins} \label{eq:Threep}.
\end{gather}

The coefficients $c_i$ of the two-point function are overlap terms given by 
\begin{gather}
    c_i=\tr[\Gamma_0 \braket{\jint_N|N_i}\braket{N_i|\jintb_N}]
\end{gather}
and coefficients $A^{i,j}_\mu$ in the three-point function are given by 
\begin{gather}
    A^{i,j}_\mu =\tr[\Gamma_k \braket{\jint_N|N_i} \braket{N_i|{\cal A}_\mu|N_j} 
    \braket{N_j|\jintb_N}],
\end{gather}
where, $\braket{N_0|{\cal A}_\mu|N_0}$ denotes the nucleon matrix element of interest in the axial case, and an analogous expression applies to the tensor three-point function. The interpolating field creates from the vacuum, i.e. $ \braket{\jint_N|N}_i\equiv\braket{\Omega|\jint_N|N_i}$,  hadron states carrying the quantum numbers of the nucleon. The states $\ket{N_i}$ for $i>0$ are excited states with energy $E_i$. To isolate the matrix elements from $A^{i,j}_\mu$, these unknown overlaps must be removed and for the ground-state matrix element, we build the ratio of the 
three-point function to the two-point function
\begin{gather}\label{eq:Ratio}
    R^\mu_A(\Gamma_k;t_s,\tins)=\frac{C_\mu(\Gamma_k,\vec{0};t_s,\tins)}{C(\Gamma_0,\vec{0};t_s)}.
\end{gather}

We note that one requires a sequential propagator in the construction of the three-point function for each projector matrix $\Gamma_k$. In the limit of large time separations, $\Delta E
(t_s-\tins)\gg 1$ and $\Delta E_1\tins\gg1$, with $\Delta E_1=E_1-E_0$,
the energy difference between the first excited state and the ground
state, this ratio gives the desired axial charge 
\begin{gather}
    R^\mu_{A}(\Gamma_k;t_s,\tins) \rightarrow g_{A}.
\end{gather}
The only non-zero contribution arises when the current is in the same direction as the projector i.e. when $\mu=k$. We then average over all spatial directions $k$.   We use the parameters
$c_0$ and $A^{0,0}$ extracted from our analysis and substitute
Eqs.~(\ref{eq:Twop}--\ref{eq:Threep}) into Eq.~(\ref{eq:Ratio})
to obtain the charges via
\beq \label{eq:ExtrRat}
 g_{A}=Z_A\,\frac{{A^{0,0}_k}(\Gamma_k)}{c_0}.
\eeq
The corresponding ratio $R_T^{\mu\nu}$
for the tensor operator yields the tensor charge,  
\beq
g_{T}=Z_T\,\epsilon^{ijk}\,\frac{{T^{0,0}_{ij}}(\Gamma_k)}{c_0},
\eeq
where $Z_A$ and $Z_T$, are renormalization constants that are computed non-perturbatively as   discussed in Sec.~\ref{sec:Renorm} and  no summation over the index $k$ is implied. In practice, one must determine the smallest value of $t_s$ at which 
excited-state contributions are adequately suppressed. The rate at which 
ground-state dominance is reached depends both on the smearing applied to 
the interpolating fields and on the energy gap separating the ground state 
from the excited states. Because the noise grows exponentially with $t_s$, 
it is crucial to establish convergence to the ground state as early as 
possible. To reliably extract the ground-state matrix elements, and thus 
determine the values of $c_0$ and $A^{0,0}$, we adopt a multi-state 
fitting strategy, as described in Sec.~\ref{sec:FitProc}.

\section{Renormalization}
\label{sec:Renorm}
The matrix elements of the axial-vector ${\cal A}_\mu$ and tensor $\mathcal{T}_{\mu\nu}$ currents are renormalized nonperturbatively in the RI$'$-MOM scheme~\cite{Martinelli:1994ty}, followed by perturbative conversion to the $\overline{\rm MS}$ scheme at the reference scale $\bar{\mu}=2$ GeV. We determine the renormalization factors for both flavor-nonsinglet ($ns$) and flavor-singlet ($s$) operators, which are needed to renormalize the isovector and isoscalar combinations. Building upon the calculation of Ref.~\cite{Alexandrou:2024ozj}, we extend the analysis to the  E112 ensemble with the smallest lattice  spacing and further improve the methodology following Refs.~\cite{Alexandrou:2025vto, Alexandrou:2026ait}. A comprehensive analysis of the nonsinglet renormalization factors will be presented in a forthcoming ETMC publication~\cite{ETMC-ren:2026}. While both singlet and nonsinglet renormalization factors are employed, we describe here  the analysis of renormalization factors for singlet operators and refer the reader to Ref.~\cite{ETMC-ren:2026} for the details of the analysis for nonsinglet ones.   Nevertheless, in this work, we use the values of both singlet and nonsinglet renormalization constants and  report their values below.\footnote{In the RI-MOM analysis of Ref.~\cite{ETMC-ren:2026} concerning  the non-singlet renormalization constants, which are  used  also here, the effect of  hadronic contaminations  on the continuum extrapolation of renormalised quark masses is discussed in detail. The singlet renormalization constants   for the nucleon charges, discussed in this work,  carry larger errors and are not sensitive to these effects. Instead, we limit the analysis to the high-momentum region to avoid effects of hadronic contaminations.}

\begin{figure*}[htbp!]
\centering
\includegraphics[width=0.49\textwidth]{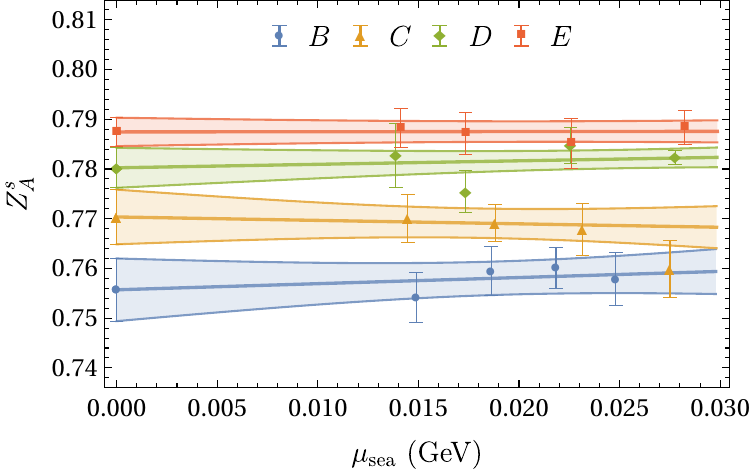} \,
\includegraphics[width=0.49\textwidth]{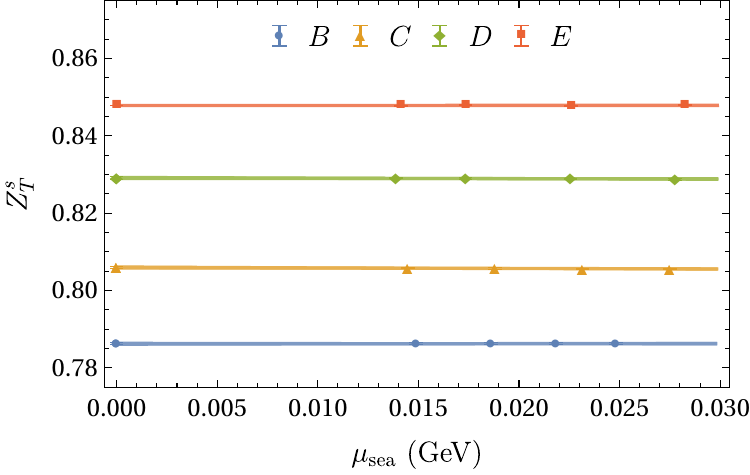}
    \caption{
    Results for the renormalization factors $Z_{A}^{\rm s}$ (left) and $Z_{T}^{\rm s}$ (right) in the RI$'$-MOM scheme at the scale ${\mu_0}^2 \simeq 21$~GeV$^2$, as a function of $\mu_{\rm sea}$. Blue circles, yellow triangles, green diamonds and red squares show, respectively, results for the $N_f=4$ ensembles at the lattice spacings of the B64, C80, D96  and E112 $N_f=2+1+1$ gauge ensembles, accompanied by the same-colored band showing the AIC-average value using constant and linear fits in $\mu_{\rm sea}$. The extrapolated values in the chiral limit are given at $\mu_{\rm sea} = 0$.}
    \label{fig:chiral_extr}
\end{figure*}
\begin{figure*}[htbp!]
\centering
\includegraphics[width=0.49\textwidth]{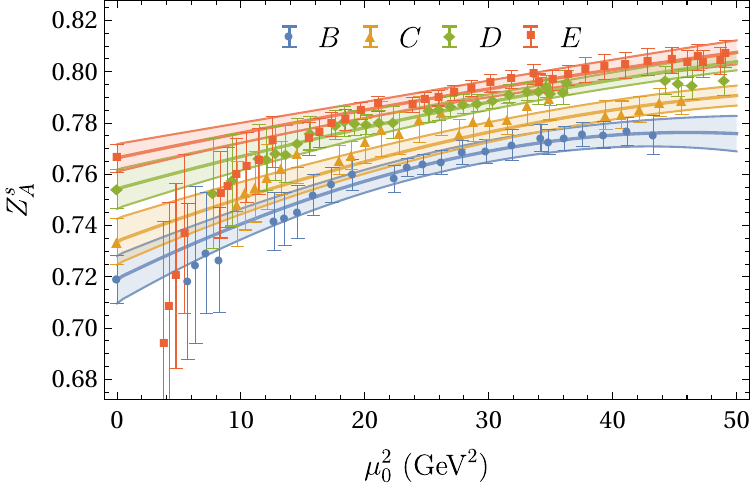} \,
\includegraphics[width=0.49\textwidth]{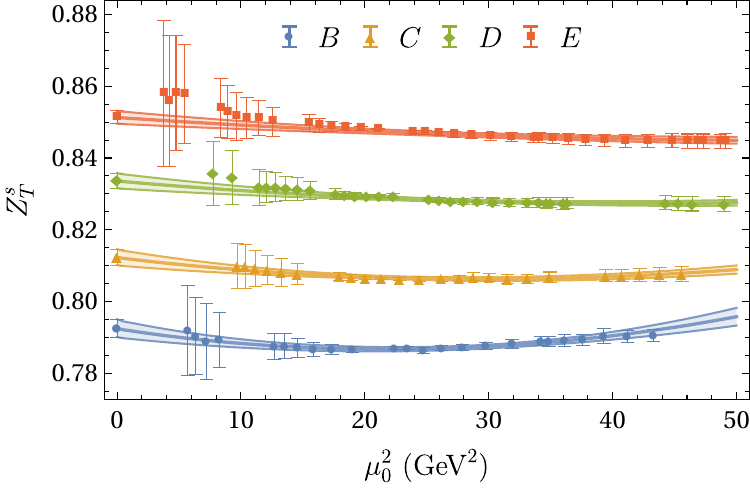} 
    \caption{Results for the renormalization factors $Z^{\rm s}_{A}$ (left), $Z^{\rm s}_T$ (right) in the RI$'$-MOM scheme at the scale $\mu_{\rm ref} = \sqrt{21}$ GeV, as a function of $\mu_0^2$. Blue circles, yellow triangles, green diamonds and red squares show, respectively, results for the $N_f=4$ ensembles with the lattice spacings of B64, C80, D96 and E112 $N_f=2+1+1$ gauge ensembles, accompanied by the same-colored band showing the AIC-average value. Polynomial fits in $a^2 \mu_0^2$ are employed in the data for several fit ranges as described in the text. The extrapolated values for each lattice spacing are given at $\mu_0^2 = 0$.}
    \label{fig:Zfactors}
\end{figure*}
The RI$'$-MOM scheme is defined on amputated vertex functions of the operators under study with external offshell quark states in Landau gauge,
\begin{equation}
    \Lambda^{\rm ns (s)}_{\mathcal{O}} (p) = \frac{a^{12}}{V} \sum_{x,y,z} e^{-i p (x-y)} \langle \psi(x) \, \mathcal{O}^{\rm ns (s)} (z) \, \bar{\psi} (y) \rangle_{\rm amp.}, 
\end{equation}
where $\mathcal{O}^{\rm ns (s)} \in \{{\cal A}_\mu^{\rm ns (s)}, \mathcal{T}_{\mu\nu}^{\rm ns (s)} \}$. Following continuum perturbation theory~\cite{Gracey:2003yr}, we project onto the ``tree-level'' structures of $\Lambda^{\rm ns (s)}_{{\cal A}_\mu}$ and $\Lambda^{\rm ns (s)}_{\mathcal{T}_{\mu\nu}}$ and impose the following conditions to extract the corresponding renormalization factors,
\begin{align}
	(Z_A^{\rm ns (s)})^{{\rm RI}'} {(Z_q^{{\rm RI}'})}^{-1}  \frac{1}{36} \sum_\mu {\rm Tr} \left[\Lambda^{\rm ns (s)}_{{\cal A}_\mu} \mathcal{P}_\mu \right] \Bigg |_{p^2 = \mu_0^2} &= 1, \\
    (Z_T^{\rm ns (s)})^{{\rm RI}'} {(Z_q^{{\rm RI}'})}^{-1}  \frac{1}{36} \sum_{\mu < \nu} {\rm Tr} \left[\Lambda^{\rm ns (s)}_{\mathcal{T}_{\mu\nu}} \mathcal{P}_{\mu \nu} \right] \Bigg |_{p^2 = \mu_0^2} &= 1,
\end{align}
where, by considering only the non-zero components of the momentum $\tilde{p}_\mu \equiv \sin(a p_\mu)$, the projectors read
\begin{align}
   \mathcal{P}_\mu &= \gamma_5 \left(\gamma_\mu - \frac{\slashed{\tilde{p}}}{4 \tilde{p}_\mu}\right), \\
   \mathcal{P}_{\mu\nu} &= \sigma_{\mu\nu} - \frac{i}{4} \left(\frac{\gamma_\mu}{\tilde{p}_\nu} - \frac{\gamma_\nu}{\tilde{p}_\mu}\right) \slashed{\tilde{p}},
\end{align}
with $\slashed{\tilde{p}} \equiv \sum_\mu  \gamma_\mu \tilde{p}_\mu$. Furthermore, $\mu_0$ denotes the RI$'$-MOM scale, and $Z_q^{{\rm RI}'}$ is the renormalization factor of the quark field defined by~\cite{ExtendedTwistedMass:2021gbo}
\begin{equation}
     Z_q^{{\rm RI}'} = \frac{1}{12} \sum_{\mu} {\rm Tr} \left[S^{-1} (p) \cdot \frac{-i \ \gamma_\mu}{4 {\tilde{p}}_\mu}\right] \Big|_{p^2 = \mu_0^2},
\end{equation}
where $S(p)$ is the quark propagator in the momentum space.  
 
The vertex function $\Lambda^{\rm ns (s)}_{\mathcal{O}} (p)$ and quark propagator $S(p)$ are calculated using momentum sources~\cite{Gockeler:1998ye} with 10-30 configurations. The singlet vertex function $\Lambda^{\rm s}_{\mathcal{O}}$ has an additional disconnected contribution, which is calculated using the stochastic methods described in Sec.~\ref{sec:stats}.

We employ gauge ensembles specifically generated for our renormalization program~\cite{Alexandrou:2024ozj}, using $N_f=4$ mass-degenerate, clover-improved twisted-mass fermions. These ensembles cover the same four lattice spacings as the $N_f=2+1+1$ physical-point ensembles used in the calculation of the matrix elements. At each lattice spacing, we consider four ensembles with different values of the twisted-mass parameter $\mu_{\rm sea}$, enabling an extrapolation to the chiral limit. The simulation parameters for the three coarsest lattice spacings are given in Ref.~\cite{Alexandrou:2024ozj}, while those for the lattice spacing of the E112 ensemble are listed in Table~\ref{tab:Nf=4 ensembles}. We find that the dependence on $\mu_{\rm sea}$ is mild, consistent with previous studies by our collaboration (see, e.g., Ref.~\cite{Alexandrou:2015sea}). We account for this dependence by performing both constant and linear fits in $\mu_{\rm sea}$ and combining the results through model averaging based on the Akaike information criterion (AIC)~\cite{Jay:2020jkz,Neil:2022joj}. A typical example is illustrated in Fig.~\ref{fig:chiral_extr} at $\mu_0^2 \simeq 21$\,GeV$^2$, for the singlet renormalization factors. 
\begin{table}[ht!]
\centering
\caption{$N_f=4$ ensembles with the lattice spacing of the E112 ensemble and their parameters used for the renormalization: $\beta=2N/g^2$, lattice volume $(L/a)^3\times(T/a)$, twisted-mass parameter $a\mu_{\rm sea}$, hopping parameter $\kappa$, and clover coefficient $c_{\rm SW}$. For the B64, C80 and D96 ensembles the corresponding parameters are given in Table~III  of Ref.~\cite{Alexandrou:2024ozj}.}
\renewcommand{\arraystretch}{1.15}
\setlength{\tabcolsep}{3pt}
\resizebox{\columnwidth}{!}{%
\begin{tabular}{c c c c c c}
\hline\hline
Ensemble & $\beta$ & $(L/a)^3 \times T/a$ & $a\mu_{\rm sea}$ & $\kappa$ & $c_{\rm SW}$ \\
\hline
cE4.035.48 & 1.960 & $48^3 \times 96$ & 0.0035 & 0.13738600 & 1.5792 \\
cE4.043.48 & 1.960 & $48^3 \times 96$ & 0.0043 & 0.13738600 & 1.5792 \\
cE4.056.48 & 1.960 & $48^3 \times 96$ & 0.0056 & 0.13738600 & 1.5792 \\
cE4.070.48 & 1.960 & $48^3 \times 96$ & 0.0070 & 0.13738600 & 1.5792 \\
\hline\hline
\end{tabular}%
}
\label{tab:Nf=4 ensembles}
\end{table}

To suppress rotational $O(4)$ breaking lattice effects, we employ spatially isotropic momenta satisfying the ``democratic'' criterion: $\sum_{\mu} p_\mu^4/(\sum_\mu p_\mu^2)^2 < 0.3$. In addition, we improve the nonperturbative results by subtracting one-loop discretization effects from both $Z_q$ and $\Lambda^{\rm ns (s)}_{\mathcal{O}}$, computed in lattice perturbation theory to all orders in the lattice spacing (see, e.g., ~\cite{Alexandrou:2015sea}).

After chiral extrapolation and artifact subtraction, the resulting renormalization factors are evolved from $\mu_0$ to a large reference scale $\mu_{\rm ref} = \sqrt{21}$ GeV using continuum perturbation theory~\cite{Chetyrkin:1997dh,Gracey:2003yr,Gracey:2022vqr,Constantinou:2016ieh}, as described in Ref.~\cite{Alexandrou:2024ozj}. Residual lattice artifacts depending on the renormalization scale $\mu_0$ are removed through polynomial fits in $a^2 \mu_0^2$. Compared to our analysis in Ref.~\cite{Alexandrou:2024ozj}, in the current work, we perform a joint fit in $a^2\mu_0^2$ using all four lattice spacings, rather than fitting each ensemble separately. To accommodate the combined dataset, higher-order terms in $a^2\mu_0^2$ are included in the fit Ansatz. Specifically, we consider 
$$f (a^2 \mu_0^2) = c_0 (a) + [c_1 (a^2 \mu_0^2) + c_2 (a^2 \mu_0^2)^2] \times \frac{R_\mathcal{O}^{{\rm RI}'} (\alpha_s (\mu^2_{\rm ref}))}{R_\mathcal{O}^{{\rm RI}'} (\alpha_s (\mu_0^2))},$$ where $R_\mathcal{O}^{{\rm RI}'} (\alpha_s (\mu^2))$ is the evolution function defined in Ref.~\cite{Alexandrou:2024ozj}. These simultaneous fits provide an improved description of the data.

We employ several fit ranges within $10 \leq \mu_0^2 \leq 40$ GeV$^2$ for the nonsiglet  $Z^{\rm ns}_A$, while we use  $22 \leq \mu_0^2 \leq 46$ GeV$^2$ for the singlet $Z^s_A$ and the tensor renormalization functions. Lower momenta are excluded from the analysis, as they suffer from significant hadronic contamination and the perturbative evolution is not reliable in this low-momentum region. The extrapolated values at $\mu_0 = 0$ from all fits are combined using model averaging based on  AIC.  Fig.~\ref{fig:Zfactors} displays the extrapolation to $\mu_0 \rightarrow 0$ for the singlet renormalization factors $Z_A^{\rm s}$ and $Z_T^{\rm s}$.

The extrapolated values at $\mu_0^2 = 0$ are converted to the $\overline{\rm MS}$ scheme at the reference scale $\bar{\mu} = 2$ GeV, using an intermediate Renormalization Group Invariant (RGI) scheme (see Ref.~\cite{Alexandrou:2024ozj}). The final values for both singlet and nonsinglet renormalization factors in the $\overline{\rm MS}$ scheme at 2 GeV and for each lattice-spacing, are reported in Table~\ref{tab:Zfactors}.
\begin{table}
\begin{center}
 \caption{Renormalization factors in the $\overline{\rm MS}$ scheme at 2 GeV for the nonsinglet and singlet axial (second and fourth rows)  and tensor  (third and fifth rows) operators used for the   B64, C80, D96 and E112 ensembles. The numbers in parentheses are the total errors, obtained by adding the statistical and systematic uncertainties in quadrature. The systematic uncertainty is determined from the AIC procedure.}
 \renewcommand{\arraystretch}{1.15}
  \begin{tabular}{lllll}
  \hline
  \hline
\ & \qquad B64 & \qquad C80 & \qquad D96 & \qquad E112 \\
\hline
\ $Z^{\rm ns}_{A}$ & \ 0.7626(10) \ & \ 0.7738(11) \ & \ 0.7851(11) \ & \ 0.7944(12) \\
\ $Z^{\rm ns}_{T}$ & \ 0.8405(71) \ & \ 0.8632(39)  \ & \ 0.8864(29)  \ & \ 0.9056(29)   \\
\ $Z^{\rm s}_{A}$ & \ 0.7579(74) \ & \ 0.7725(73)\ & \ 0.7911(50) \ & \ 0.7993(37) \\
\ $Z^{\rm s}_{T}$ & \ 0.8433(40) \ & \ 0.8645(40) \ & \ 0.8872(40)\ & \ 0.9058(36)  \\
\hline
\hline
  \end{tabular}
  \label{tab:Zfactors}
  \end{center}
\end{table}

\section{Fitting Procedure}\label{sec:FitProc}

\subsection{The two-step approach}
To extract the ground-state matrix elements, we employ a two-step fitting strategy. 

In the first step, we fit the effective mass,
\begin{gather}\label{eq:EffMass}
    m_N^{\text{eff}}= \ln \left( \frac{ C(\Gamma_0,\vec{0};t_s)}{C(\Gamma_0,\vec{0};t_s+1)} \right)
\end{gather}
constructed from the highest-statistics two-point functions, using Eq.~(\ref{eq:EffMass}) and Eq.~(\ref{eq:Twop}) with $i=0,1$
\begin{gather}\label{eq:EM_twost}
    m_N^{\text{eff}}= m_0 +\ln \left( \frac{1+r_1e^{-\Delta E_1\,t}}{1+r_1e^{-\Delta E_1(t+1)}} \right),
\end{gather}
which we refer to as a \textit{two-state} fit.  We look for convergence as a function of the smallest time separation $t^{\rm low}$ used in the fit. The extracted values of  the ground-state energy $m_0 = E_0$ and the energy gap $ \Delta E_1=E_1-E_0$, together with the corresponding ratio of overlap factors $r_1=c_1/c_0$ entering Eq.~(\ref{eq:EM_twost})  are subsequently used as fixed inputs in the analysis of the ratio.

In the second step, we construct the ratio of Eq.~(\ref{eq:Ratio}) by dividing each three-point function by the corresponding two-point function computed with the same statistics in order to take advantage of the correlations between them. The ratio is then fitted using either a two-state fit or to a constant when there is no detectable contamination due to excited states. We refer to the latter as a \textit{plateau} fit. For the two-state fit we use 
\begin{equation}\label{eq:RatioFit}
\begin{aligned}
R(t_s,\tins) = g\Bigg[
&\frac{
1+a_{10}\Big(
e^{-\Delta E_1\tins}
+e^{-\Delta E_1(t_s-\tins)}
\Big)
}{
1+r_1e^{-\Delta E_1t_s}
}
\\[4pt]
&+\frac{
a_{11}e^{-\Delta E_1t_s}
}{
1+r_1e^{-\Delta E_1t_s}
}
\Bigg].
\end{aligned}
\end{equation}
where $g=A^{0,0}/c_{0}$ the charge and $a_{ij}=A^{i,j}/A^{0,0}$, the ratios of overlap factors of the three-point function. As mentioned, we fix the values of $m_0, \Delta E_1,r_1$ from the effective mass fit in the first step.  

For the isovector axial case, our fit incorporates the temporal component 
of the three-point function carrying one unit of momentum transfer, 
$C_0(\Gamma_k, \frac{2\pi}{L}\hat{k}; t_s, \tins)$, where $L$ denotes the spatial extent of the lattice and $\hat{k}$ is a unit vector along the $k$-th spatial direction, together with the corresponding two-point function with one unit of momentum. Including in the analysis the three-point function of the temporal axial current  is motivated by chiral perturbation theory~\cite{Bar:2018xyi,Bar:2016uoj}, which predicts an enhancement of the overlap between  a nucleon and a pion-nucleon state, namely the   $A^{0,1}_\mu$ and $A^{1,0}_\mu$ coefficients. In our setup, the temporal component of the axial operator comes at no extra computational cost, and because it displays a strong sensitivity to the excited state, it enables a more precise determination of the excited state energy of the lowest $\pi N$ state and hence a better extraction of the desired ground-state matrix element.

Specifically, we form a ratio using $C_0(\Gamma_k, \frac{2\pi}{L}\hat{k}; t_s, \tins)$ and the matched statistics two-point functions at $\vec{q}{=}0$ and $\vec{q}{=}\frac{2\pi}{L}\hat{k}$, and the effective energy at maximum statistics using Eq.~(\ref{eq:EffMass}) with $\vec{q}{=}\frac{2\pi}{L}\hat{k}$. The two-state fit function for the ratio of the temporal component we use is
\begin{equation}\label{eq:A0Ratio}
\begin{aligned}
R_{0}(t_s,\tins)
&= \frac{C_0( \vec{q}, t_s, \tins)}{C(\vec{0},t_s)}
\\
&\quad\times
\sqrt{
\frac{
C(\vec{q},t_s-\tins)
C(\vec{0},\tins)
C(\vec{0},t_s)
}{
C(\vec{0},t_s-\tins)
C(\vec{q},\tins)
C(\vec{q},t_s)
}
},
\end{aligned}
\end{equation}
where the three-point function is
\begin{equation}
\begin{aligned}
C_0( \vec{q}, t_s, \tins)
&= A^{0,0}_0\,
e^{-m_0(t_s-\tins)-E_0(\vec{q})\tins}
\\
&\quad\times\Big[
1+b_{10}e^{-\Delta E_1(t_s-\tins)}
\\
&\qquad
+b_{01}e^{-\Delta E_1(\vec{q})\tins}
\\
&\qquad
+b_{11}
e^{-\Delta E_1(t_s-\tins)-\Delta E_1(\vec{q})\tins}
\Big].
\end{aligned}
\end{equation}
The two-point functions entering the ratio are
\begin{equation}
C(\vec{0},t)
= c_0e^{-m_0t}
\left(1+r_{1}e^{-\Delta E_1t}\right),
\end{equation}
and
\begin{equation}
C(\vec{q},t)
= c_0(\vec{q})e^{-E_0(\vec{q})t}
\left(1+r_{1}(\vec{q})e^{-\Delta E_1(\vec{q})t}\right),
\end{equation}
with
\begin{equation}
E_0(\vec{q})=\sqrt{m_0^2+\vec{q}^{\,2}},
\qquad |\vec{q}|=\frac{2\pi}{L}.
\end{equation}

In this case, the first-excited-state energy gap $\Delta E_1$ of the three-point function at zero momentum transfer is shared with the corresponding one for the three-point function for temporal axial current. For the two-point function, $\Delta E_1$ is fixed from the effective mass fit. 

In all cases, we compare our approach to the results from the \textit{summation} method where we retain only the ground state. Specifically, we perform a linear fit in $t_s$ using the following summed ratio
\begin{align}
     S^{\mu}_{A}(\Gamma_k;t_s) = &\sum^{t_s-a}_{\tins=a}   R^\mu_A(\Gamma_k;t_s,\tins) \nonumber\\
     \xrightarrow{\Delta Et_{s}\gg1}&~c+ g_A t_s,
\end{align}
for the axial charge and similarly for $S^{\mu \nu}_{T}(\Gamma_k;t_s)$ for the tensor.

\subsection{Extraction of ground state matrix
elements}
\label{sec:results}

\begin{figure}[htbp!]
	\includegraphics[width=\columnwidth]{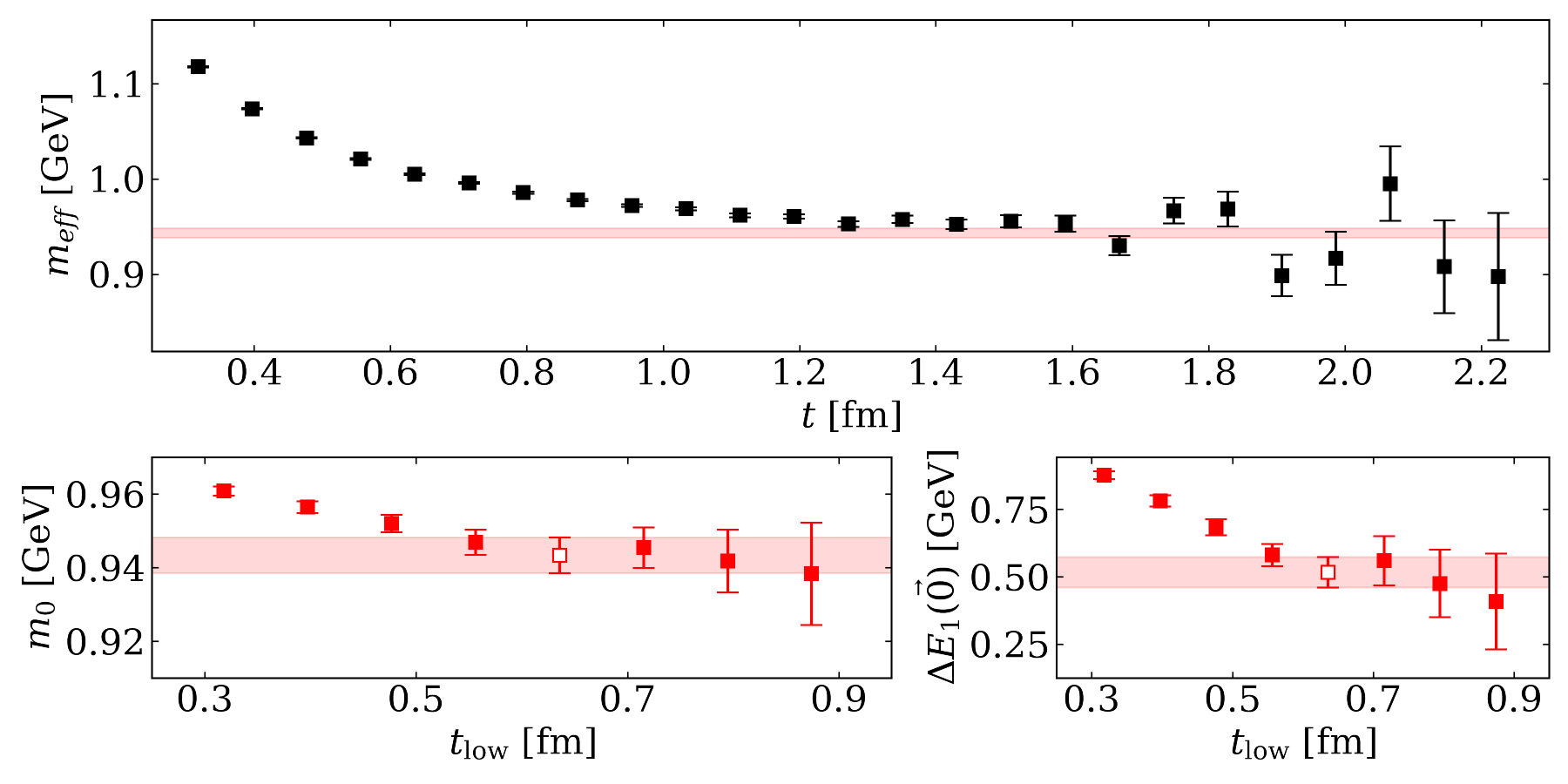}\\
	\vspace*{-0.3cm}
    \caption{Two-state fit results for the nucleon effective mass, for the B64 ensemble. The upper panel shows results on the effective nucleon mass $m_{\rm eff}$ as a function of the
    source-sink time separation $t$. The lower-left panel shows results on the nucleon mass $m_0$ extracted from two-state fits as a function of the minimum fit time $t_\text{low}$. The lower-right panel shows the corresponding first excited-state energy gap $\Delta E_1$. Open symbols indicate the
    selected fit results, while the red bands show the value from the selected two-state fit.}
	\vspace*{-0.4cm}
    \label{fig:eff_mass}
\end{figure}

A detailed description of the  analysis of the nucleon effective mass using the approach explained in Sec.~\ref{sec:FitProc} can be found in Ref.~\cite{Alexandrou:2026oks}. We use the same procedure here to obtain the needed parameters for the first step in our approach. We illustrate our results in Fig.~\ref{fig:eff_mass} for the B64 ensemble. We observe that after $t^{\rm low}_{m_{\text{eff}}}\sim 0.6$~fm the extracted values of  the ground-state energy $m_0 = E_0$ and $ \Delta E_1=E_1-E_0$, are stable. This is valid  for all gauge ensembles. We thus fix $t_{m_{\rm eff}}^{\rm low}\approx0.6$~fm.

\begin{figure}[htbp!]
	\includegraphics[width=\columnwidth]{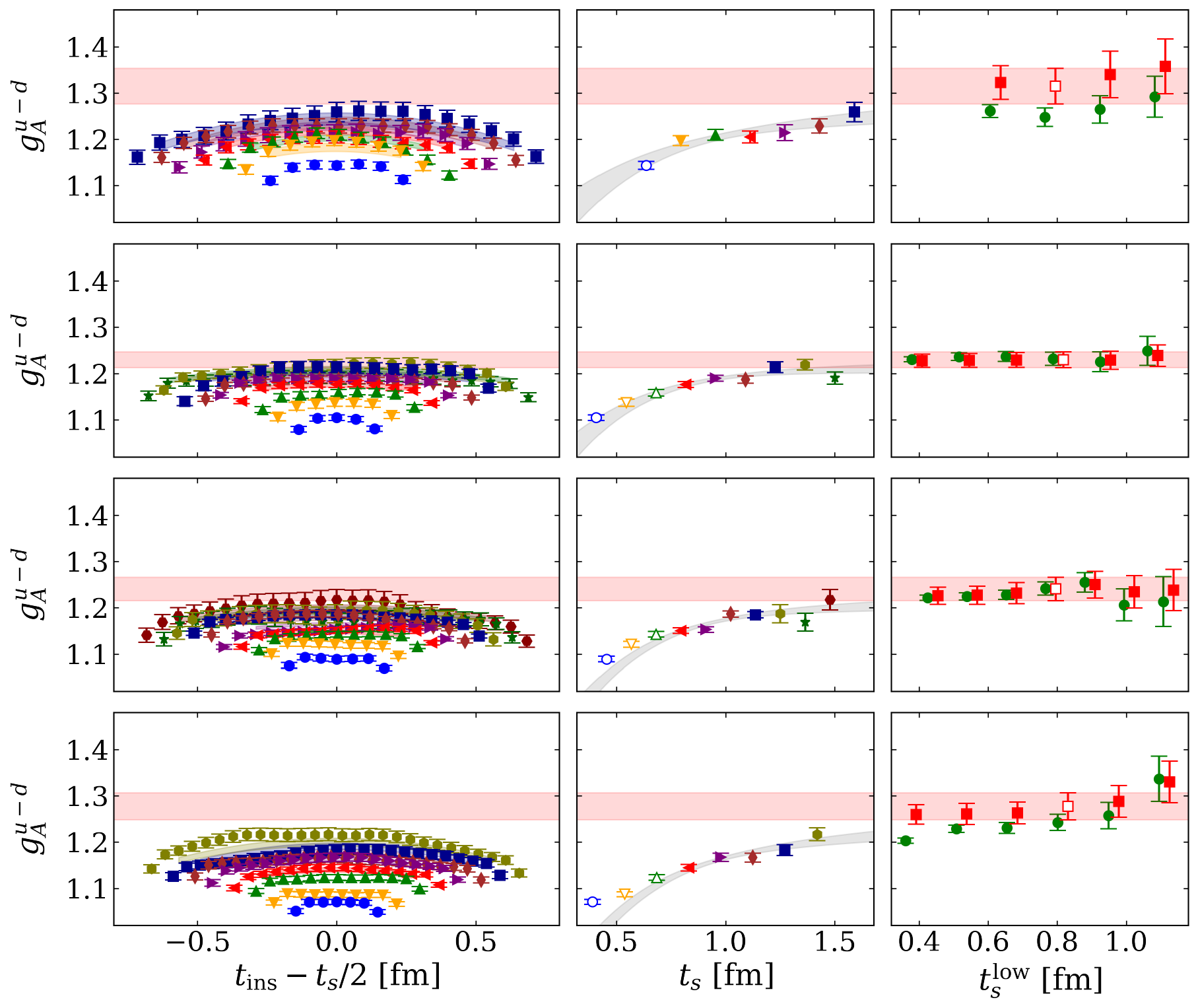}\\
	\vspace*{-0.3cm}
    \caption{Results for the isovector axial charge $\gaumd$. From top to bottom
we show results for the B64, C80, D96, and E112 ensembles. Left panels show the ratio as a function of $t_{\text{ins}}-t_s/2$ for different source-sink time separations $t_s$. The middle panels show the midpoint value of the ratio  as a function of $t_s$ using the same symbols for each time separation $t_s$ as in the left panels. The right panels show the values extracted from two-state fits (red squares) as a function of the smallest value of $t_s$, $t_s^{\text{low}}$, used in the fits. The open symbols indicate the selected fit. The horizontal bands correspond to the chosen fit result.
In the left and middle panels, the gray bands show the
resulting curves using the   selected two-state fit parameters. The filled symbols in the middle panels represent the time separations used in the best  fit that is selected. Results from the summation method (green circles) are also shown for comparison.}
	\vspace*{-0.4cm}
    \label{fig:gA_umd}
\end{figure}

\begin{figure}[htbp!]
	\includegraphics[width=\columnwidth]{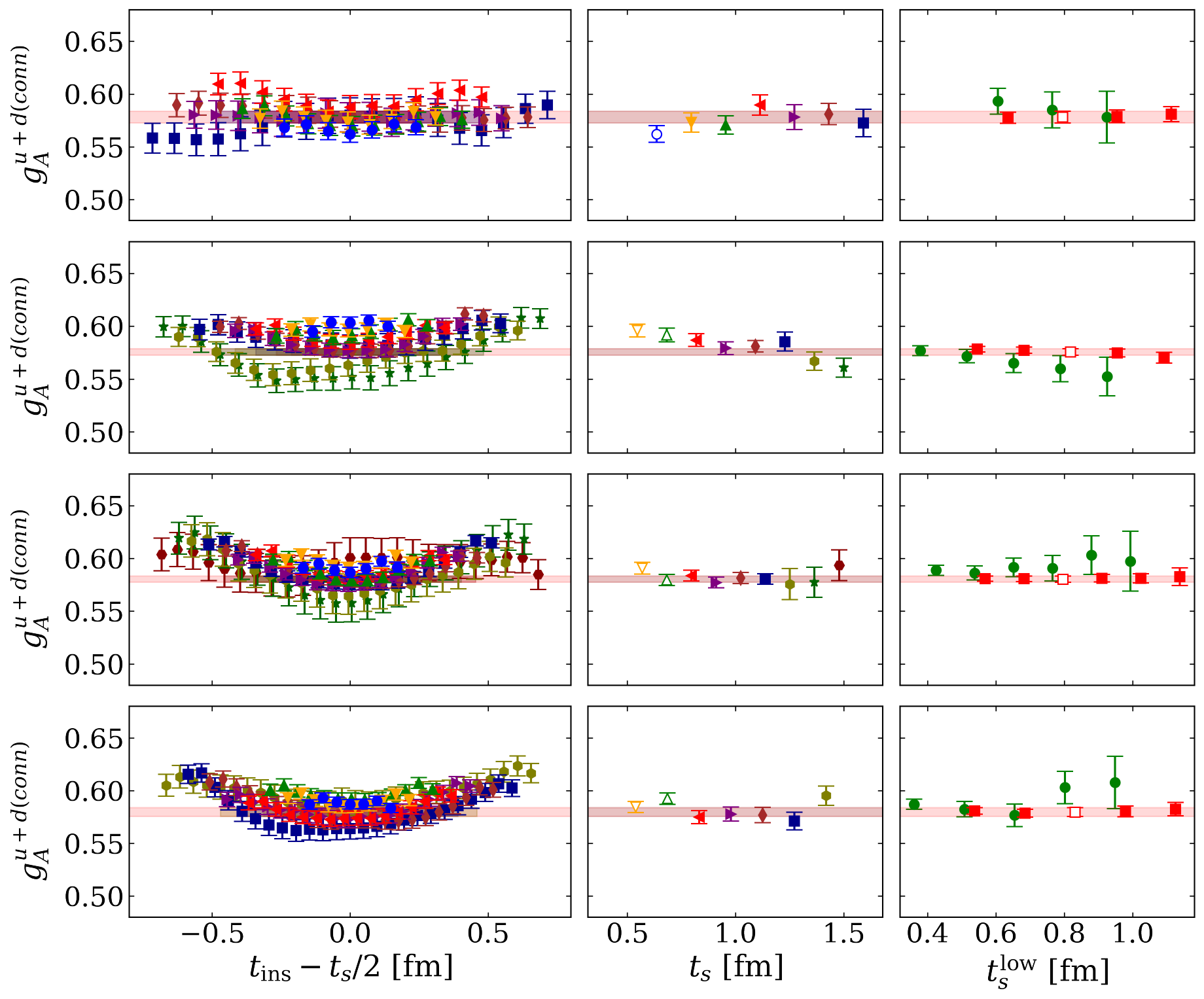}\\
	\vspace*{-0.3cm}
    \caption{Results for the ratios with the three-point ratio of the connected  isoscalar axial current (left panels). The middle panels show the values extracted from plateau fits to each source-sink separation $t_s$  individually as a function of $t_s$. The right panels show the results from plateau fits (red squares) including all
    source-sink separations satisfying $t_s \ge t_s^{\text{low}}$ , where the open
    symbols indicate the selected fit. The band corresponds to
    the chosen fit. Results from the summation method (green
    circles) are also shown for comparison. The rest of the notation is the same as in Fig.~\ref{fig:gA_umd}.}
	\vspace*{-0.4cm}
    \label{fig:gA_upd}
\end{figure}

\begin{figure}[htbp!]
	\includegraphics[width=\columnwidth]{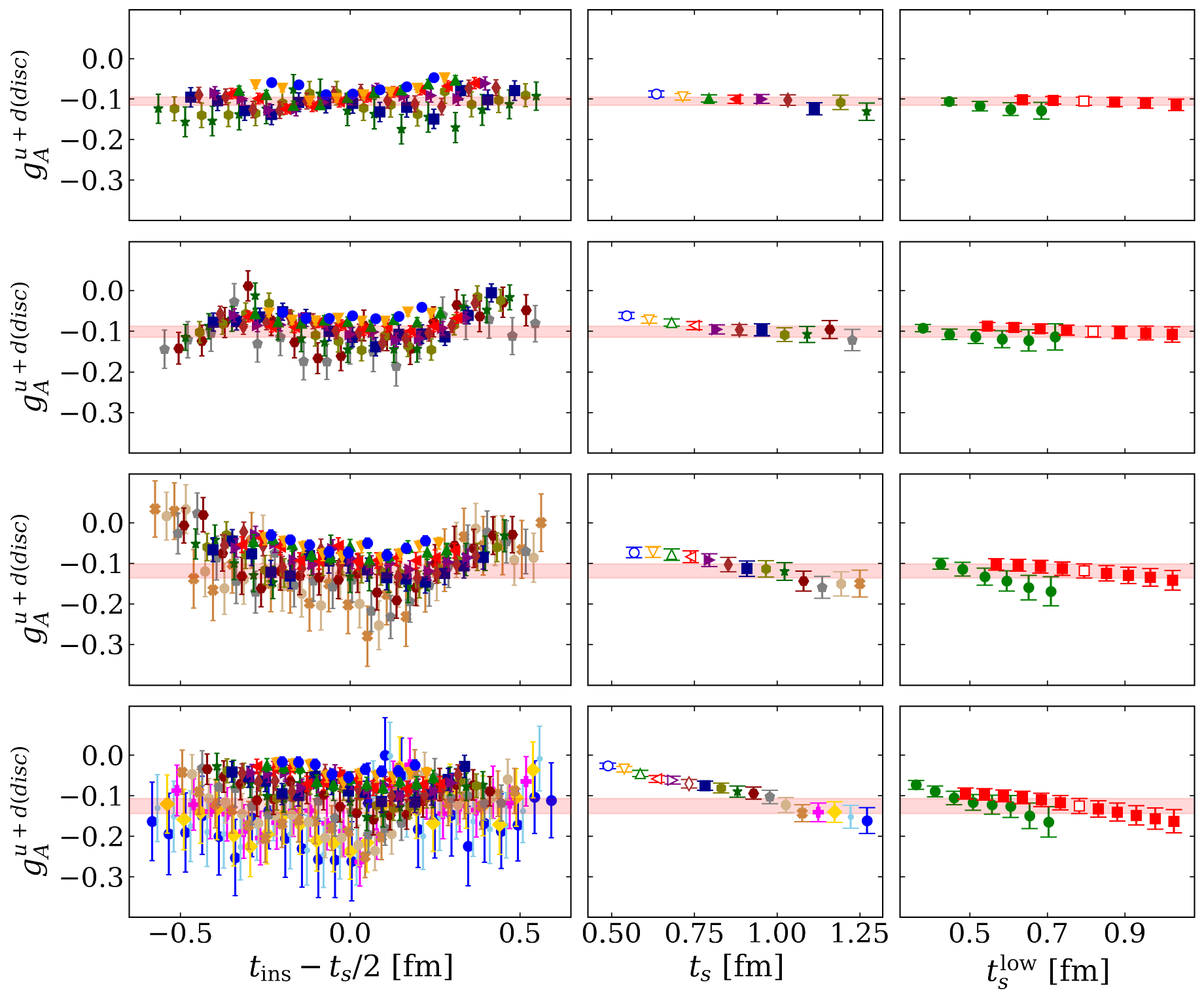}\\
	\vspace*{-0.3cm}
    \caption{Results for the disconnected contribution of the isoscalar axial charge $\gaupd$. The notation is the same as in Fig.~\ref{fig:gA_upd}.}
	\vspace*{-0.4cm}
    \label{fig:gA_upd_disc}
\end{figure}

\begin{figure}[htbp!]
	\includegraphics[width=\columnwidth]{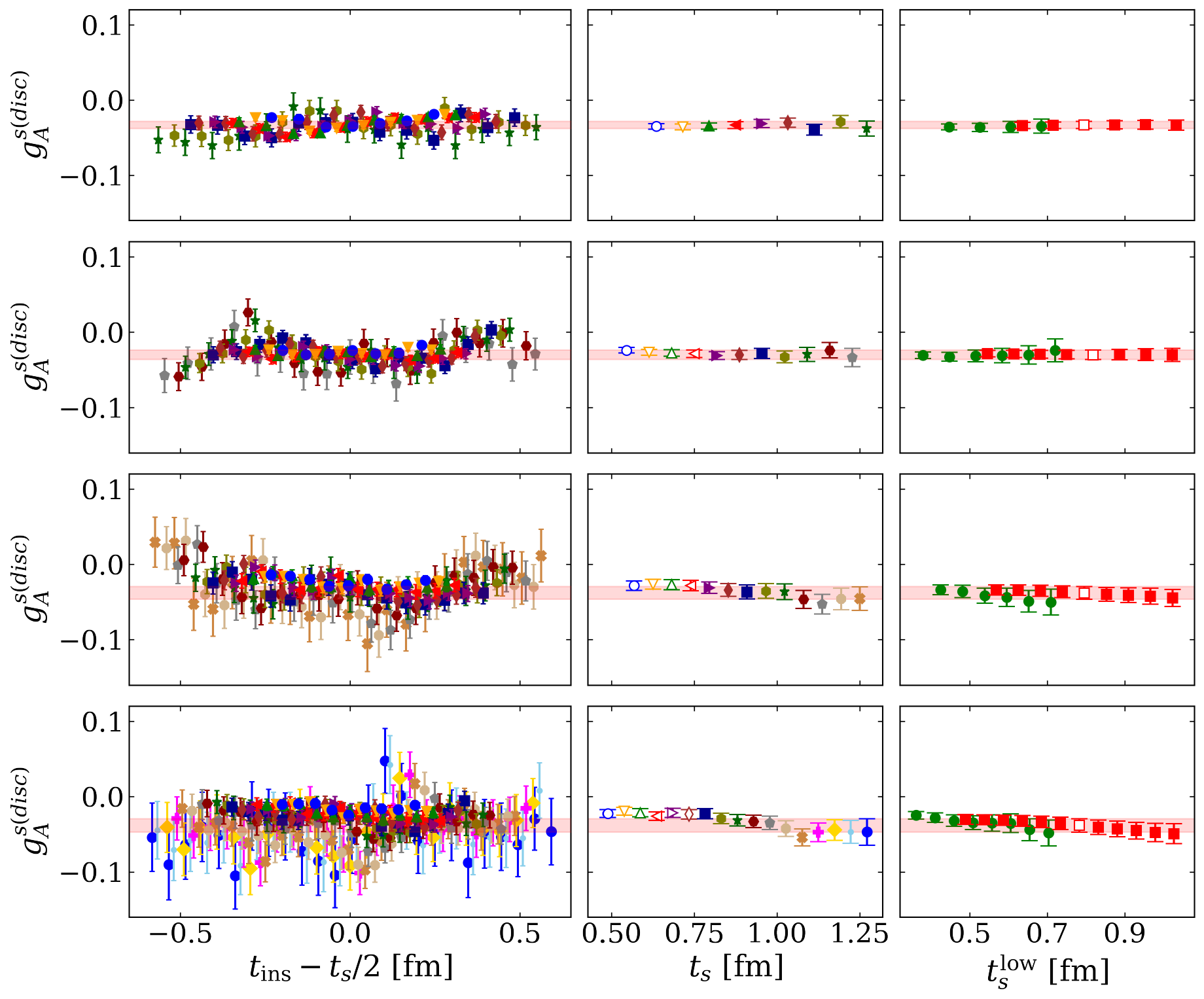}\\
	\vspace*{-0.3cm}
    \caption{Results for the strange contribution to the axial charge. The notation is the same as in Fig.~\ref{fig:gA_upd}.}
	\vspace*{-0.4cm}
    \label{fig:gA_s_disc}
\end{figure}

\begin{figure}[htbp!]
	\includegraphics[width=\columnwidth]{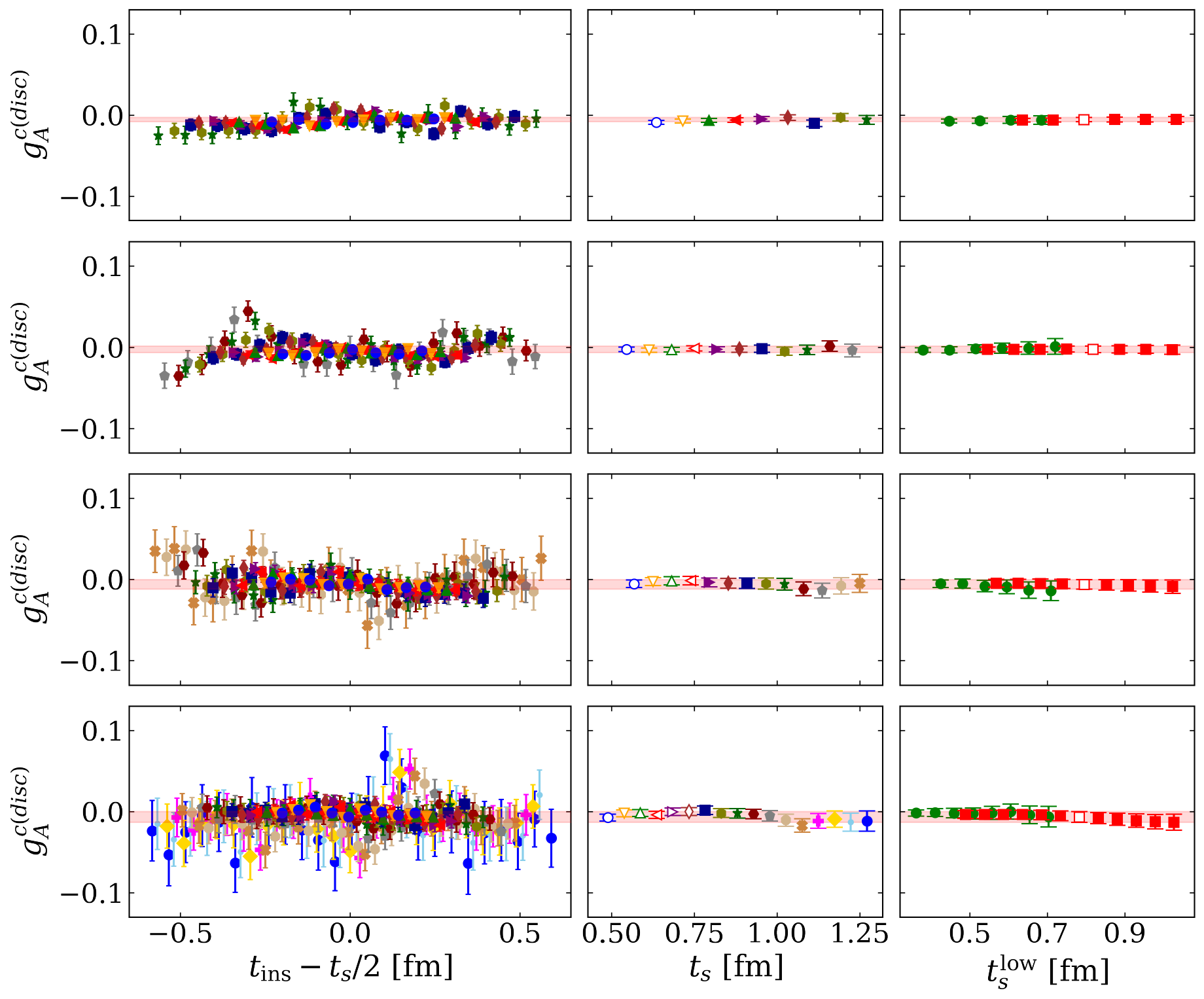}\\
	\vspace*{-0.3cm}
    \caption{Results for the charm contribution to the axial charge. The notation is the same as in Fig.~\ref{fig:gA_upd}.}
	\vspace*{-0.4cm}
    \label{fig:gA_c_disc}
\end{figure}

Results on the fits carried out for the extraction of the nucleon ground state matrix element for the isovector, isoscalar connected and disconnected, strange and charm axial charges are shown, respectively, in Figs.~\ref{fig:gA_umd}-\ref{fig:gA_c_disc} for the four ensembles analyzed here. As mentioned previously, we obtain the desired charge from the matrix element of the appropriate current at zero momentum transfer. 

As can be seen in Fig.~\ref{fig:gA_umd}, the ratio with numerator the connected three-point function, from which the isovector combination to the axial charge is determined, exhibits a considerable dependence on $\tins$, which indicates sizable
excited state contributions. For this case, we therefore perform
two-state fits, following the specific fitting approach of
Sec.~\ref{sec:FitProc}. We perform different fits within the range $t_\text{ins}^\text{cut} \leq t_\text{ins} \leq t_s -t_\text{ins}^\text{cut}$, eliminating time slices from the source and sink to identify convergence to a time-independent value. We find that taking a time distance of  $t_\text{ins}^\text{cut} \approx 0.2$ fm from the source and sink is an appropriate choice. We also vary the source-sink time separations satisfying $t_s \ge t_s^{\text{low}}$. We find that $t_s^\text{low} \approx 0.8$ fm  is a suitable value for all ensembles.

In Fig.~\ref{fig:gA_upd}, we show results on the ratio from which the connected contribution of the isoscalar axial charge is determined. We observe that, apart from some time-dependence near the source and sink,  it does not exhibit strong dependence on  $\tins$ and thus we perform a plateau fit. This can be seen from fitting  to a constant at each time separation in order to probe the dependence on $t_s$, which is shown in the middle panels. We use $t_\text{ins}^\text{cut} \approx 0.2$ fm to remove the time-dependence near the sink and source. We observe that the extracted values  for $t_s \ge 0.8$ fm are consistent. Thus, we perform a final plateau fit to all values of $t_s$, starting from $t_s^{\text{low}} \approx 0.8$ fm, while using the same $t_\text{ins}^\text{cut}$. As can be seen, the values are stable as we vary  $t_s^{\text{low}}$,
and they agree with the obtained values when fitting each $t_s$ separately, for our picked range. We also perform fits using the summation method as an alternative procedure. This gives statistically noisier, but compatible results with the plateau fit results.

The ratios constructed from disconnected three-point functions, namely for the isoscalar and the strange and charm quarks, are shown in Figs.~\ref{fig:gA_upd_disc}, ~\ref{fig:gA_s_disc} and ~\ref{fig:gA_c_disc}. Since no significant excited state effects are observed,   we perform plateau fits using the same ranges as for the ratios with the isovector three-point function, namely, we take $t_\text{ins}^\text{cut} \approx 0.2$~fm and $t_s^{\text{low}} \approx 0.8$~fm. 

\begin{figure}[htbp!]
	\includegraphics[width=\columnwidth]{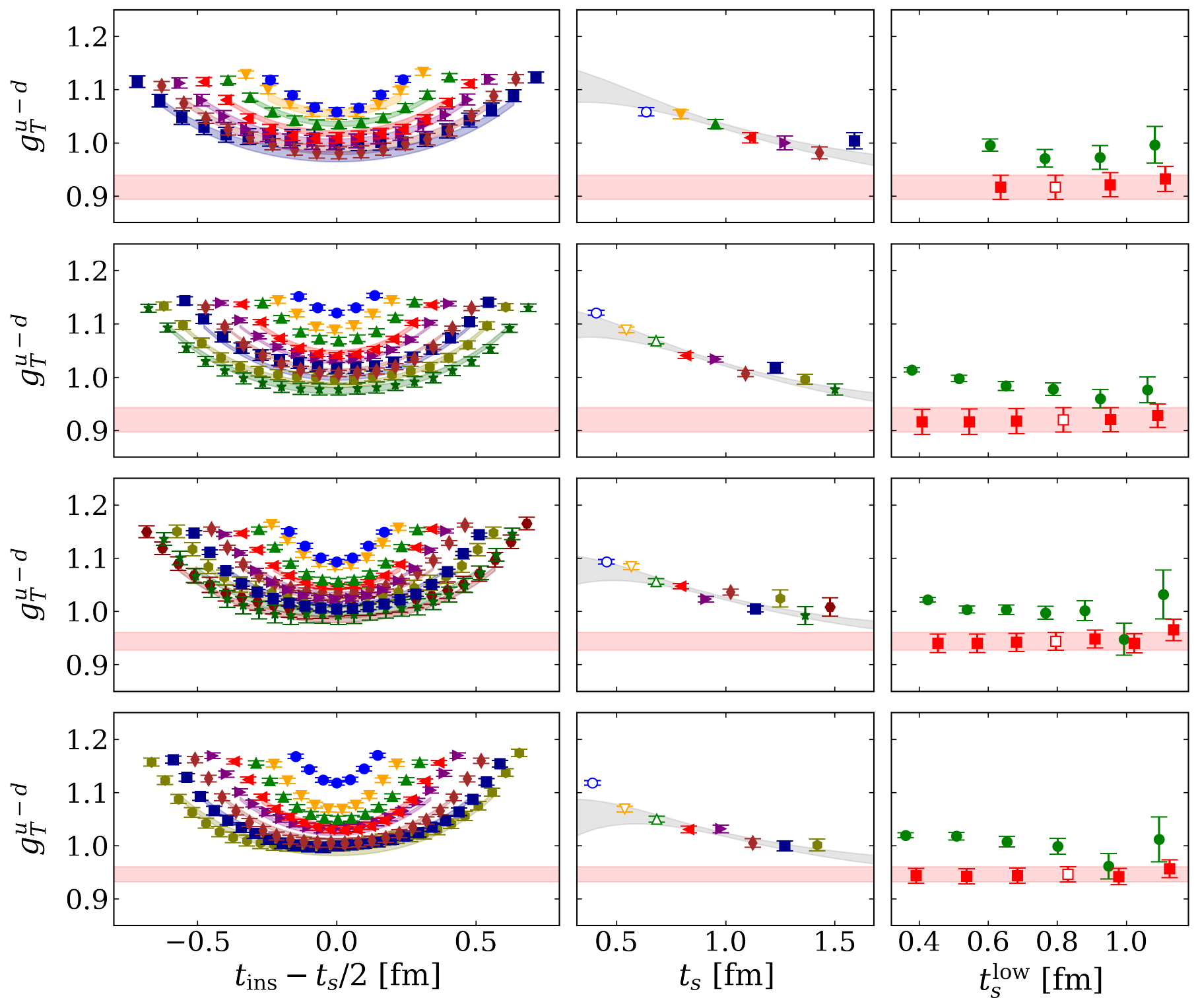}\\
	\vspace*{-0.3cm}
    \caption{Results for the isovector tensor charge $g_T^{u-d}$. The notation is the same as in Fig.~\ref{fig:gA_umd}.}
	\vspace*{-0.4cm}
    \label{fig:gT_umd}
\end{figure}

\begin{figure}[htbp!]
	\includegraphics[width=\columnwidth]{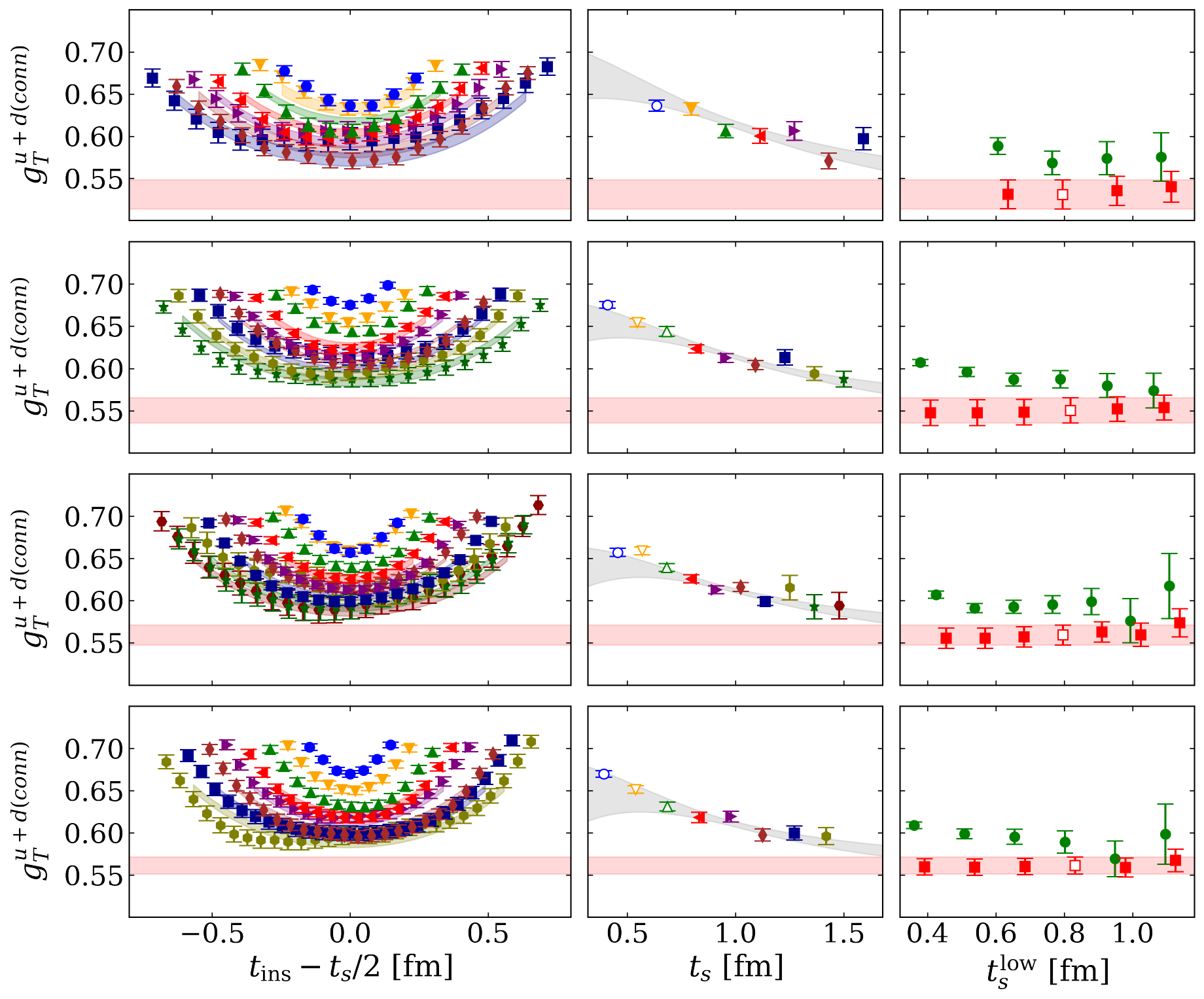}\\
	\vspace*{-0.3cm}
    \caption{Results for the connected contribution of the isoscalar tensor charge $g_T^{u+d}$. The notation is the same as in Fig.~\ref{fig:gA_umd}.}
	\vspace*{-0.4cm}
    \label{fig:gT_upd}
\end{figure}

\begin{figure}[htbp!]
	\includegraphics[width=\columnwidth]{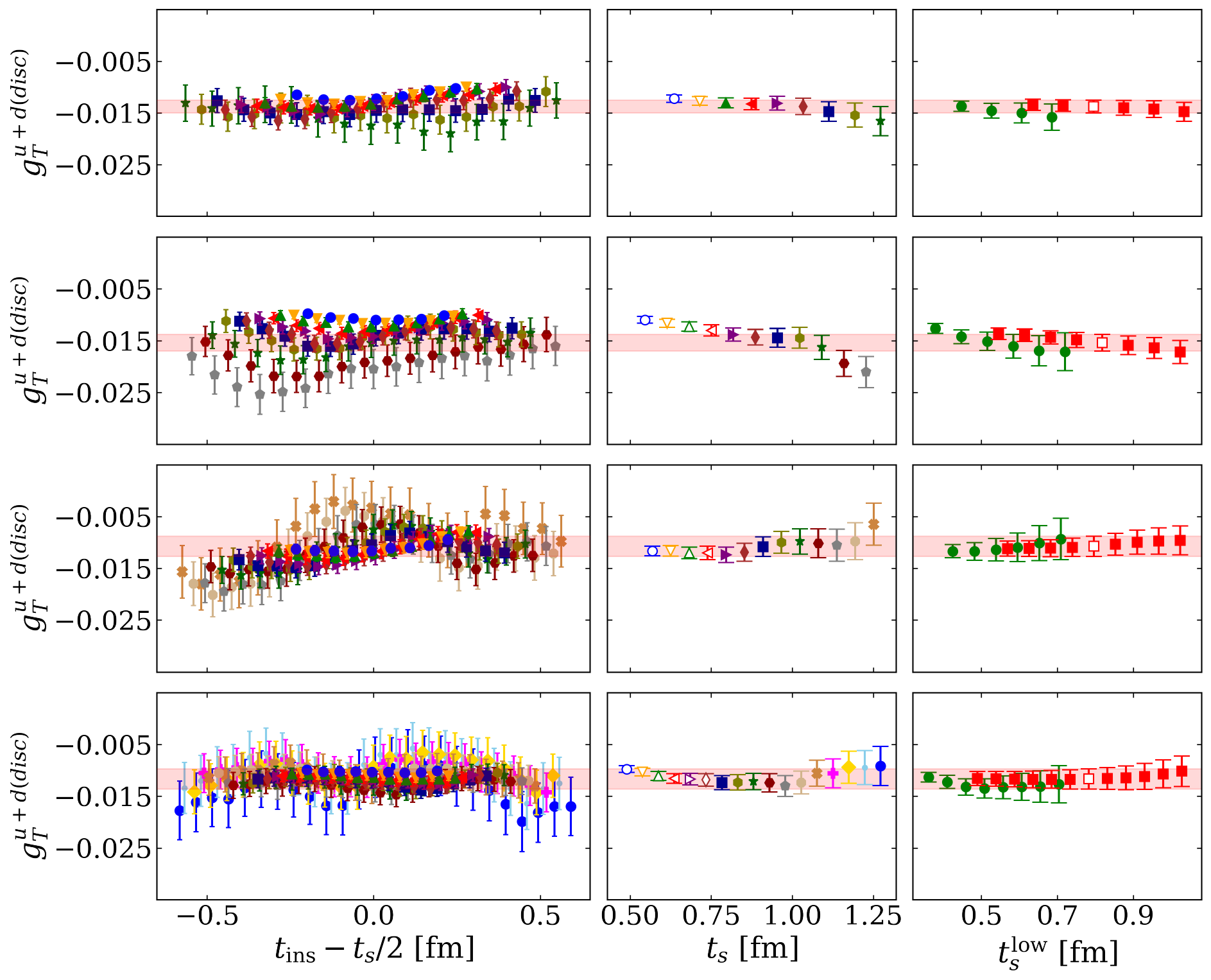}\\
	\vspace*{-0.3cm}
    \caption{Results for the disconnected contribution of the isoscalar tensor charge $\gaupd$. The notation is the same as in Fig.~\ref{fig:gA_upd}.}
	\vspace*{-0.4cm}
    \label{fig:gT_upd_disc}
\end{figure}

\begin{figure}[htbp!]
	\includegraphics[width=\columnwidth]{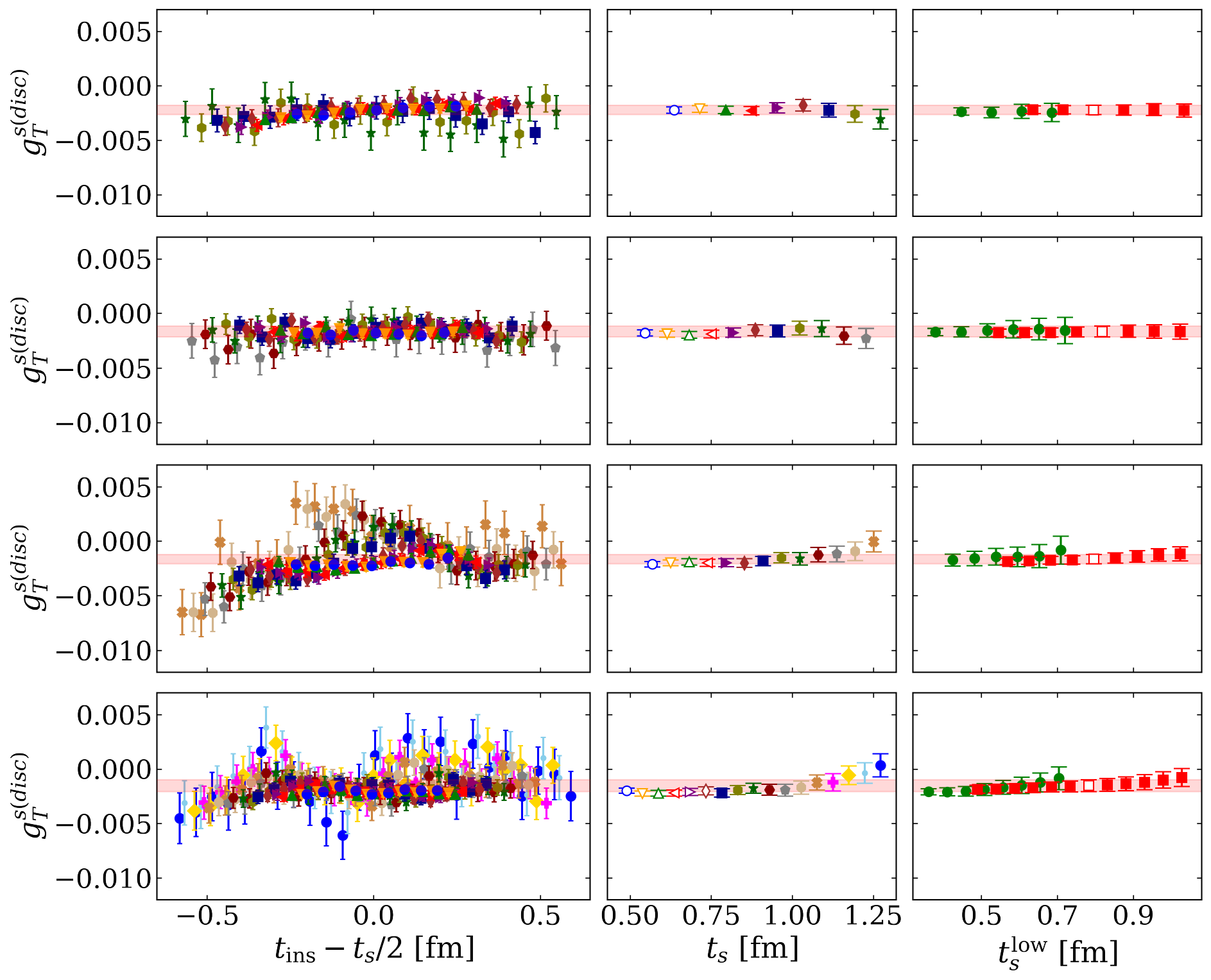}\\
	\vspace*{-0.3cm}
    \caption{Results for the strange contribution to the tensor charge. The notation is the same as in Fig.~\ref{fig:gA_upd}.}
	\vspace*{-0.4cm}
    \label{fig:gT_s_disc}
\end{figure}

\begin{figure}[htbp!]
	\includegraphics[width=\columnwidth]{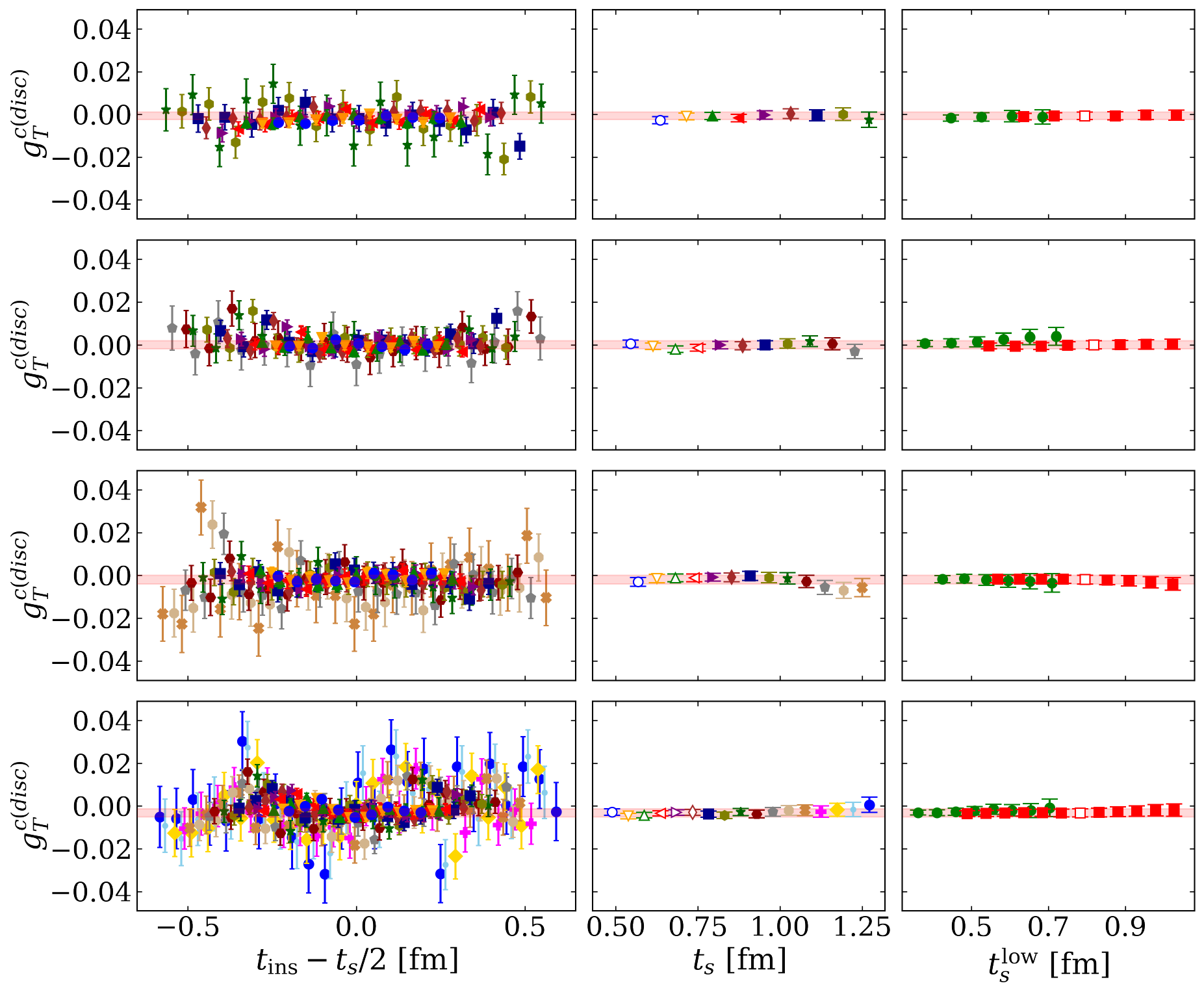}\\
	\vspace*{-0.3cm}
    \caption{Results for the charm contribution to the tensor charge. The notation is the same as in Fig.~\ref{fig:gA_upd}.}
	\vspace*{-0.4cm}
    \label{fig:gT_c_disc}
\end{figure}

In Figs.~\ref{fig:gT_umd} and~\ref{fig:gT_upd} we show our results from the analysis of the ratio for the connected isovector and isoscalar contributions to the tensor charge respectively. We observe strong excited state contamination in both quantities, thus we perform two-state fits, using $t_\text{ins}^\text{cut} \approx 0.2$~fm and $t_s^{\text{low}} \approx 0.8$~fm for all ensembles. 

The disconnected contributions to the tensor charge are shown in  Figs.~\ref{fig:gT_upd_disc}-~\ref{fig:gT_c_disc}. We observe no detectable excited state contamination within the accuracy of our results. Thus, we perform plateau fits using the same $t_s^{\text{low}}$ and $t_\text{ins}^\text{cut}$.

\subsection{Continuum extrapolation}\label{sec:FitProcC}

In the twisted mass formulation  and after renormalization, the leading lattice discretization effect are of $\ord(a^2)$. Thus, we perform linear extrapolations in $a^2$ in order to extract continuum results. In the cases where we observe no statistically significant $a^2$-dependence in our data, so we also include constant fits in our analysis and take the model average using the Akaike information criterion (AIC). Namely, we use two linear fits in $a^2$, either
using all four ensembles or omitting the coarser ensemble, and three constant fits using all ensembles or the two and three finer ones. A strong dependence on the lattice spacing will result in a model average favoring the linear fits, while a mild $a^2$ dependence will lead to a model average favoring the constant fits.

\begin{figure}[htbp!]
	\includegraphics[width=\linewidth]{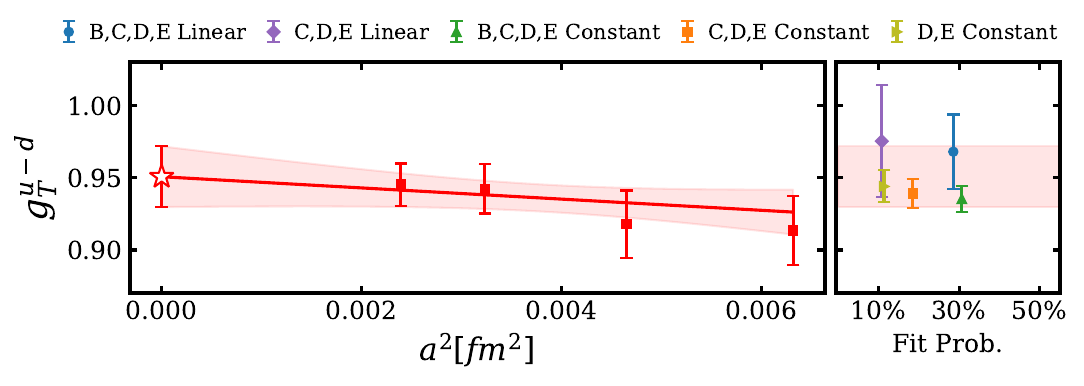}\\
	\vspace*{-0.3cm}
	\caption{Continuum extrapolation of the isovector tensor charge $g_T^{u-d}$. In the left panel the filled points are the renormalized values for each of the four ensembles. The red band and the star show the AIC model averaged results using linear and constant fits. In the right panel, we show the weights of each extrapolation model as a $\%$-probability. The red band is the AIC model averaged result at the continuum. }
	\vspace*{-0.4cm}
    \label{fig:gT_umd_extrap}
\end{figure}

\begin{figure*}[htbp!]
    \centering
    \includegraphics[
        height=6cm
    ]{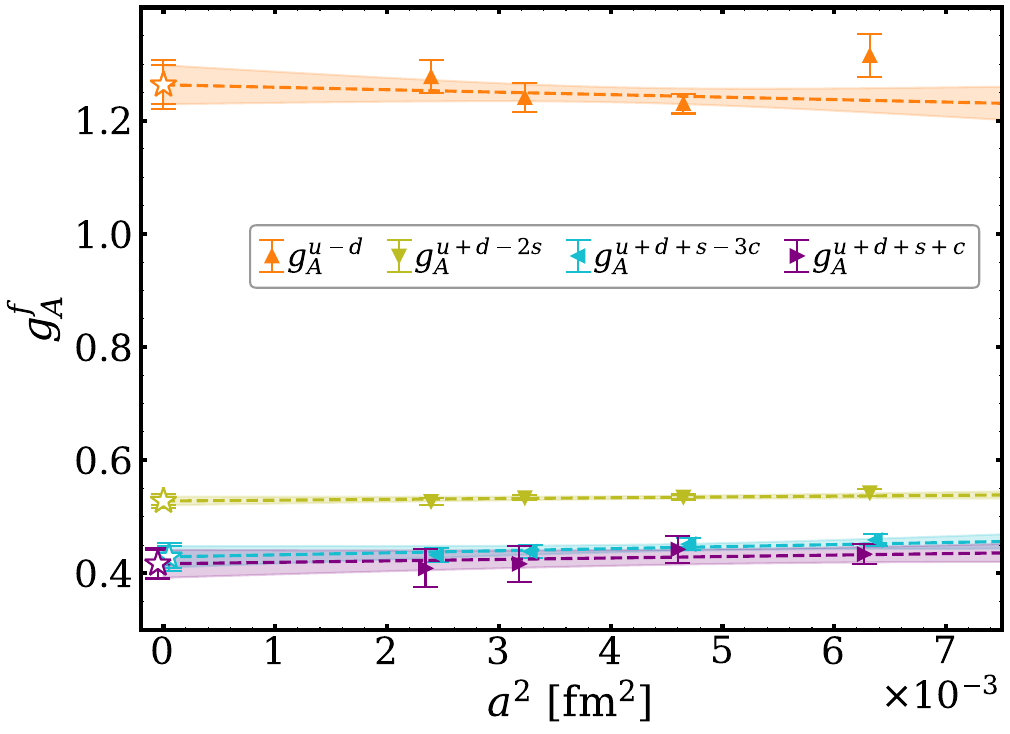}%
    \includegraphics[
        height=6cm,
        trim=80 0 0 0,
        clip
    ]{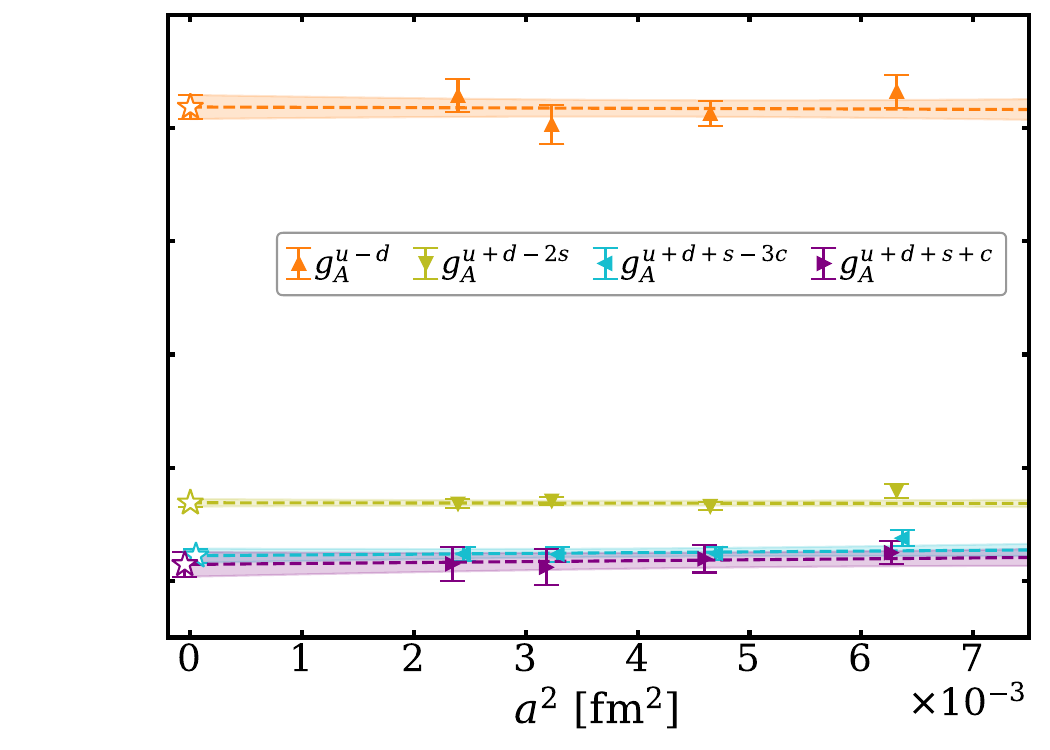}
    \caption{Continuum extrapolations of the non-singlet and the singlet quark combinations for  the axial charge. The left panel shows the results using the analysis as described in Sec.~\ref{sec:FitProcC}, with the stars showing the values at $a=0$, where the inner error is the statistical and the outer the total uncertainty after adding the systematic error in quadrature. In the right column we show the results of the analysis from which the systematic error is obtained, as explained in Sec.~\ref{sec:Syst}.}
    \label{fig:gA_su4}
\end{figure*}

\begin{figure*}[htbp!]
    \centering
    \includegraphics[
        height=6cm
    ]{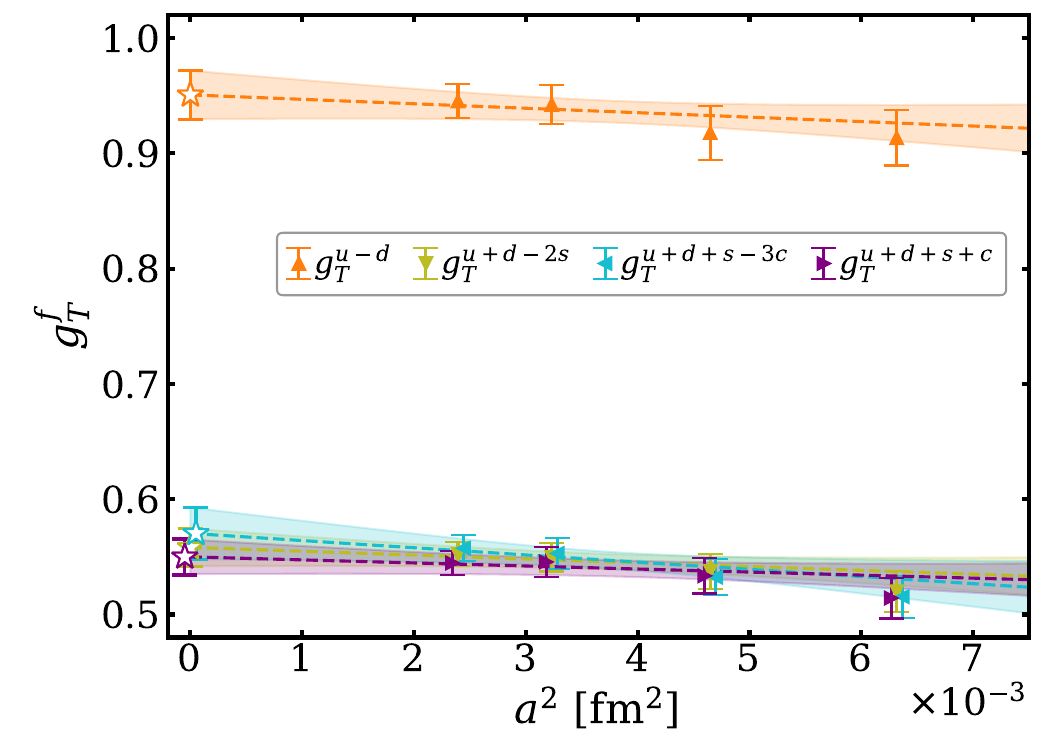}%
    \includegraphics[
        height=6cm,
        trim=80 0 0 0,
        clip
    ]{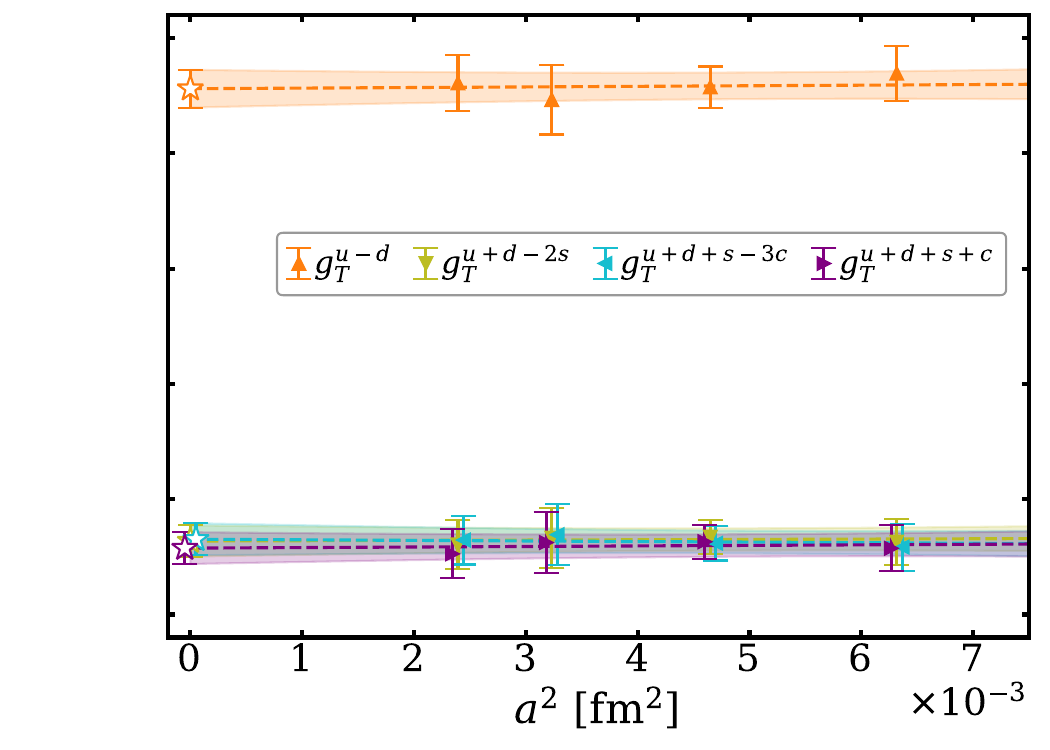}
    \caption{Continuum extrapolations of the non-singlet and the singlet quark combinations for the tensor charge. The notation is the same as in Fig.~\ref{fig:gA_su4}.}
    \label{fig:gT_su4}
\end{figure*}

We form the non-singlet combinations $u{-}d$, $u{+}d{-}2s$, $u{+}d{+}s{-}3c$ and the singlet $u{+}d{+}s{+}c$. We renormalize multiplicatively with the factors of Table~\ref{tab:Zfactors}, using $Z_A^{\rm ns}$ and $Z_T^{\rm ns}$ for the non-singlets and $Z_A^{\rm s}$ and $Z_T^{\rm s}$ for the singlet. An example extrapolation for the isovector tensor charge $g_T^{u-d}$ is shown in Fig.~\ref{fig:gT_umd_extrap}, along with the weights of the five models used. In the left panels of Figs.~\ref{fig:gA_su4} and~\ref{fig:gT_su4} we show the extrapolations for all combinations, for the axial and tensor charge respectively. 

\begin{figure*}[htbp!]
    \centering
    \includegraphics[
        height=6cm
    ]{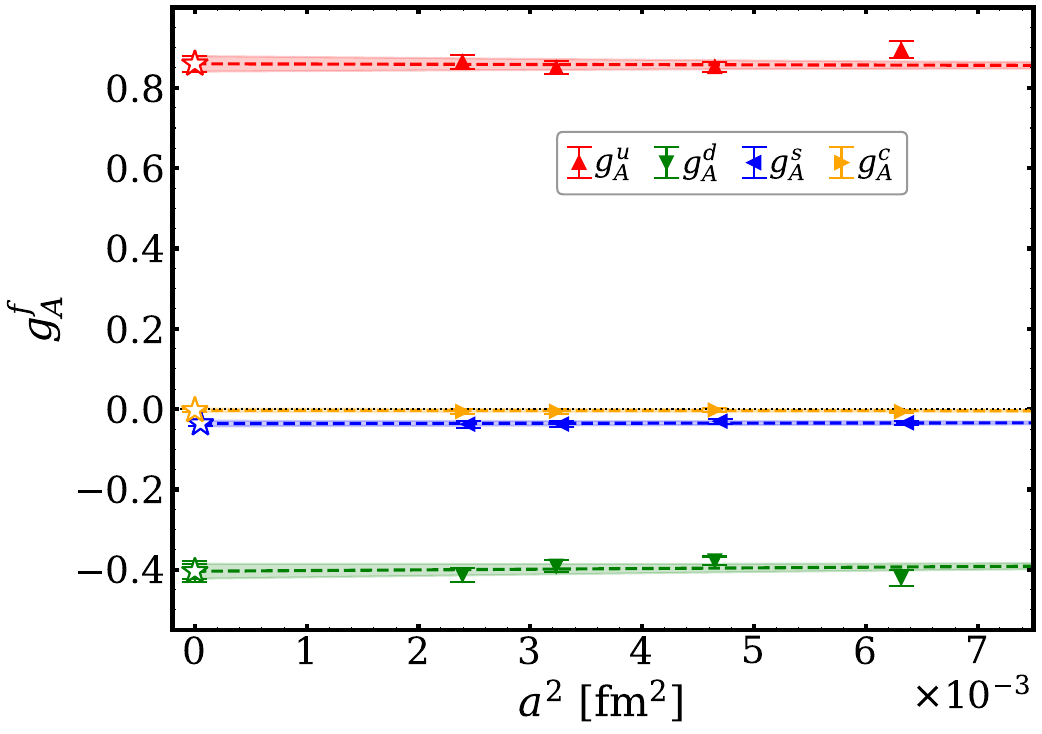}%
    \includegraphics[
        height=6cm,
        trim=80 0 0 0,
        clip
    ]{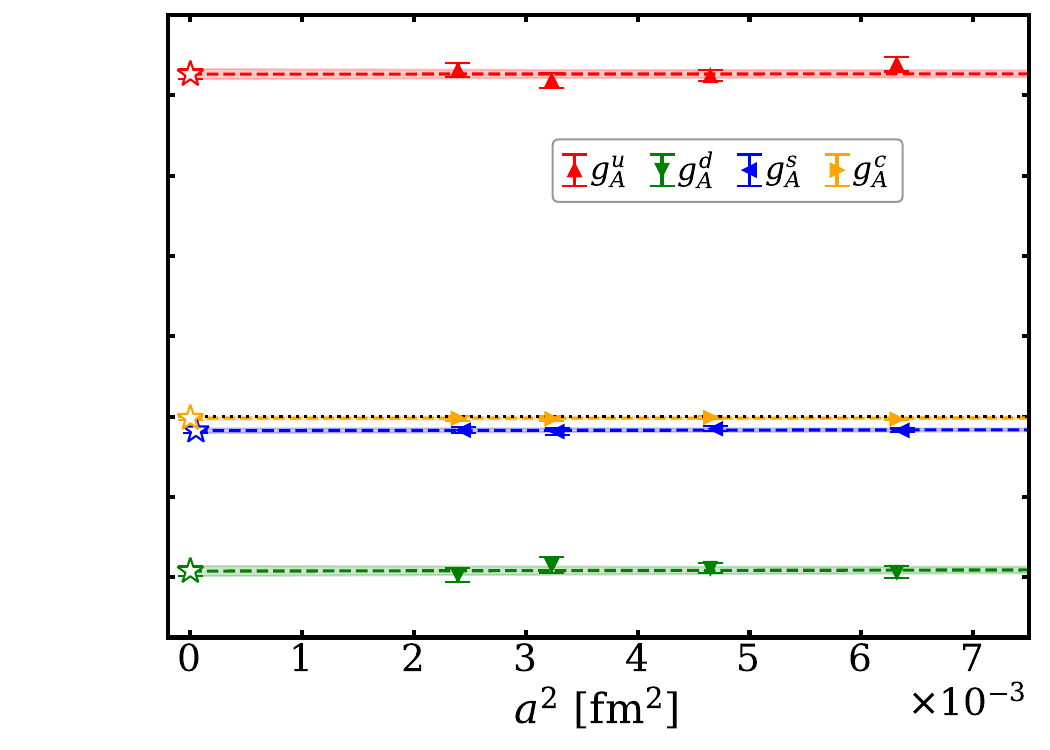}
    \caption{Continuum results for the flavor decomposition of the axial charge. The stars show the values at $a=0$, obtained using the continuum values of the non-singlet and singlet combinations. The filled symbols are the decomposed values per ensemble, used only for visualization. The bands show the linear combination of the fit bands of the respective isovector and isoscalar combinations in Fig.~\ref{fig:gA_su4}. The left and right panels show results from the analyses as described in Fig.~\ref{fig:gA_su4}.}
    \label{fig:gA_flavs}
\end{figure*}

\begin{figure*}[htbp!]
    \centering
    \includegraphics[
        height=6cm
    ]{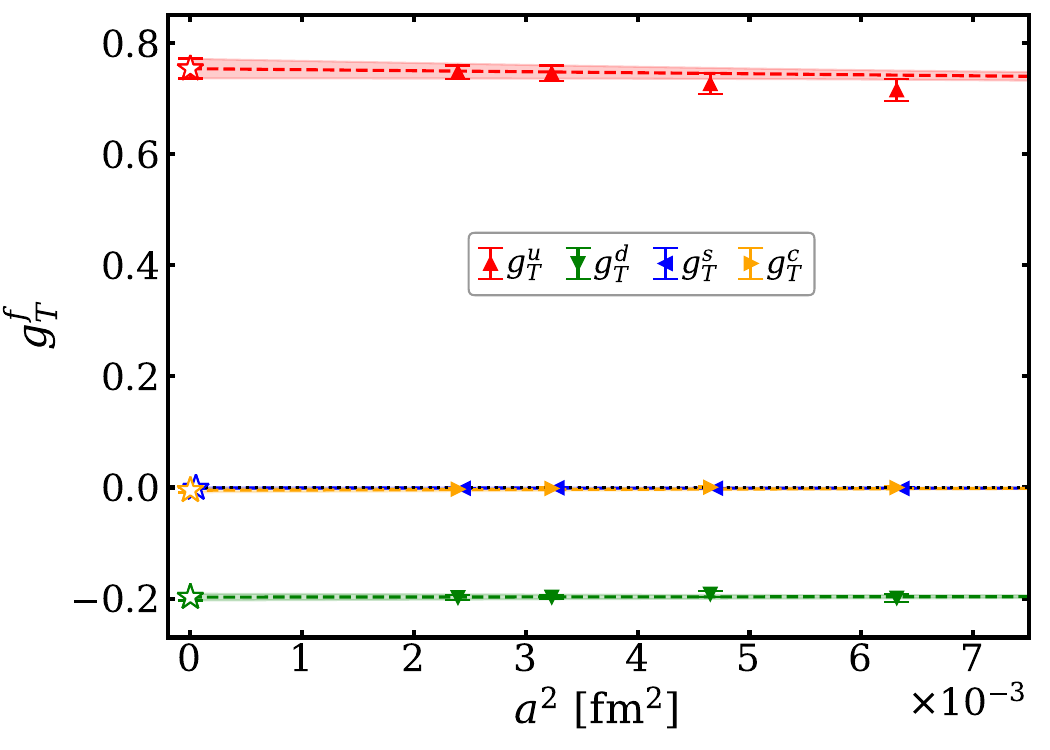}%
    \includegraphics[
        height=6cm,
        trim=80 0 0 0,
        clip
    ]{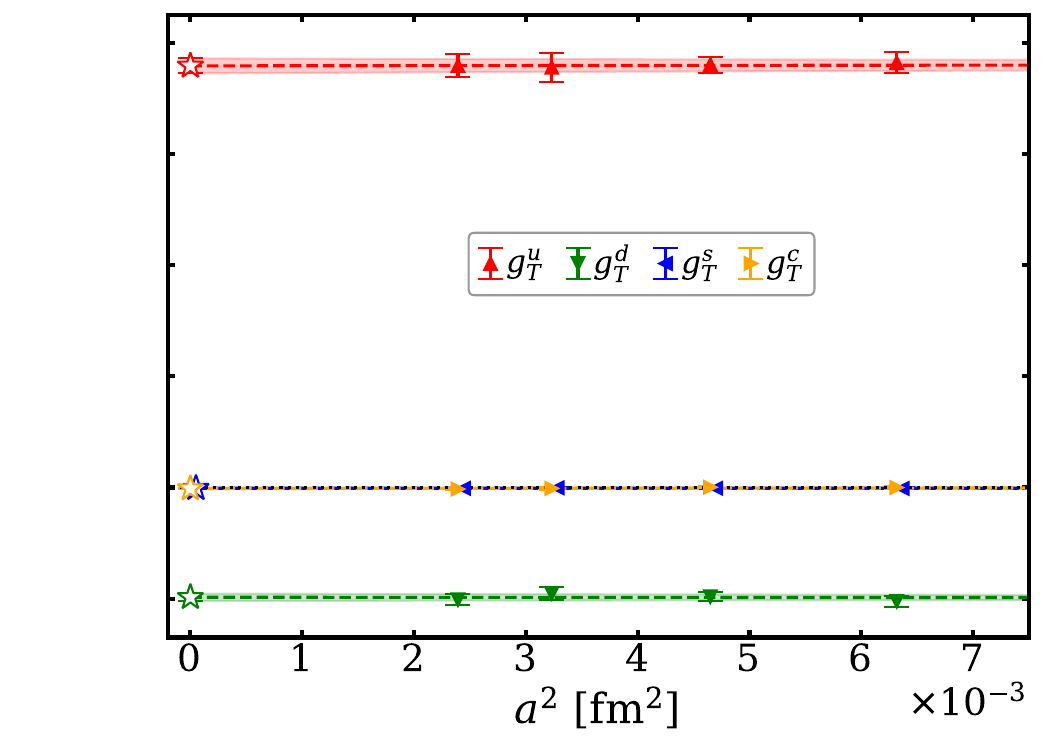}
    \caption{Continuum results for the flavor decomposition of the tensor charge. The notation is the same as in Fig.~\ref{fig:gA_flavs}.}
    \label{fig:gT_flavs}
\end{figure*}

Using results for the three non-singlet and the singlet
quark combinations at the continuum limit, we reconstruct the individual quark-flavor contributions $u,~d~,s,~c$. In the left panels of Figs.~\ref{fig:gA_flavs} and~\ref{fig:gT_flavs}, we show the results for the flavor decomposition, for the axial and tensor charges, respectively.

\subsection{Systematic uncertainties}\label{sec:Syst}

We investigate  systematics arising from the analysis of
excited states, where no AIC average was done. For the analysis of  connected contributions we will use the summation method to estimate the systematic error. We take the value at  $t_s^\text{low} \approx 1.0$~fm  where the results from two-state fits show a general agreement. For the connected isoscalar axial case $\gaupd$, where  convergence to the two-state results occurs earlier due to the absence of excited state contamination,  we take the value at $t_s^\text{low} \approx 0.6$~fm. For the disconnected contributions, instead of performing a single plateau fit to all selected values of $t_s>t_s^{\rm low}$ simultaneously, we fit to a constant at each $t_s$ value and take the weighted average after convergence, which is typically observed for the
same values of $t_s^\text{low}$ as the common fit to all separations. The results from this varied analysis are shown in the right panels of Figs.~\ref{fig:gA_su4}-~\ref{fig:gT_flavs}. 

The systematic uncertainty on the continuum value is computed by taking the absolute difference between the central value of the preferred analysis and the one when changing the procedure as described above, after continuum extrapolation.

\begin{table*}[htbp!]
	\centering
	\caption{Values for the axial non-singlet and singlet  combinations (top) and the extracted single flavor charges (bottom) for each ensemble and in the
          continuum limit, using the model average strategy described
          in the text. For the continuum values, the first error is statistical and the second the systematic due to the excited state analysis.}
	\label{tbl:gA}
	{\normalsize
		\renewcommand{\arraystretch}{1.2}
		\renewcommand{\tabcolsep}{2.0pt} 
		\begin{tabular*}{\textwidth}{@{\extracolsep{\fill}}cccccc@{}}
        \hline \hline
               & $g_A^{u-d}$   & $g_A^{u+d}$  & $g_A^{u+d-2s}$  & $g_A^{u+d+s-3c}$ & $g_A^{u+d+s+c}$  \\ \hline
        B64    & 1.315(39)     &  0.473(12)     & 0.5426(62)      & 0.459(10)       & 0.434(18)  \\ 
        C80    & 1.230(17)     &  0.475(15)     & 0.5350(44)      & 0.452(11)       & 0.442(24)  \\ 
        D96    & 1.241(25)     &  0.461(18)     & 0.5337(43)      & 0.439(12)       & 0.417(32)  \\
        E112   & 1.277(29)     &  0.454(19)     & 0.5273(58)      & 0.432(13)       & 0.409(34)  \\ \hline
        $a= 0$ & 1.263(34)(26) &  0.458(16)(1)  & 0.5280(77)(101) & 0.429(19)(15)   & 0.417(25)(12)  \\ \hline \hline
		\end{tabular*}
        \\[0.1cm]
		\begin{tabular*}{\textwidth}{@{\extracolsep{\fill}}ccccc@{}}
        \hline \hline
               & $g_A^{u}$     & $g_A^{d}$      & $g_A^{s}$        & $g_A^{c}$          \\ \hline
        B64    & 0.895(21)     & -0.420(19)     & -0.0341(48)      & -0.0062(28)  \\ 
        C80    & 0.852(12)     & -0.378(11)     & -0.0301(64)      & -0.0023(42)  \\ 
        D96    & 0.850(16)     & -0.391(15)     & -0.0371(85)      & -0.0055(60)  \\
        E112   & 0.865(18)     & -0.413(17)     & -0.0375(88)      & -0.0059(65)  \\ \hline
        $a= 0$ & 0.860(19)(7)  & -0.404(19)(19) & -0.0360(70)(10)  & -0.0030(46)(8)  \\ \hline \hline
		\end{tabular*}
	}
	\vspace*{-0.2cm}
\end{table*}

\begin{table*}[htbp!]
	\centering
	\caption{Values for the tensor non-singlet and singlet  combinations (top) and the extracted single flavor charges (bottom) for each ensemble and in the
          continuum limit. The notation is the same as in Table~\ref{tbl:gA}.}
	\label{tbl:gT}
	{\normalsize
		\renewcommand{\arraystretch}{1.2}
		\renewcommand{\tabcolsep}{2.0pt} 
		\begin{tabular*}{\textwidth}{@{\extracolsep{\fill}}cccccc@{}}
        \hline \hline
          & $g_T^{u-d}$       & $g_T^{u+d}$   & $g_T^{u+d-2s}$& $g_T^{u+d+s-3c}$& $g_T^{u+d+s+c}$  \\ \hline
        B64    & 0.913(24)    & 0.517(17)     & 0.520(18)     & 0.515(18)      & 0.514(18)  \\ 
        C80    & 0.918(23)    & 0.535(15)     & 0.538(15)     & 0.532(16)      & 0.533(15)  \\ 
        D96    & 0.942(17)    & 0.548(12)     & 0.550(12)     & 0.553(13)      & 0.546(13)  \\
        E112   & 0.945(15)    & 0.549(10)     & 0.552(10)     & 0.557(11)      & 0.545(11)  \\ \hline
        $a= 0$ & 0.951(21)(5) & 0.556(16)(5)  & 0.558(16)(6)  & 0.570(22)(5)   & 0.550(15)(8)  \\ \hline \hline
		\end{tabular*}
        \\[0.1cm]
		\begin{tabular*}{\textwidth}{@{\extracolsep{\fill}}ccccc@{}}
        \hline \hline
               & $g_T^{u}$     & $g_T^{d}$      & $g_T^{s}$        & $g_T^{c}$          \\ \hline
        B64    & 0.715(19)    & -0.1986(70)    & -0.0018(17)    & -0.0003(21)  \\ 
        C80    & 0.726(19)    & -0.1916(58)    & -0.0014(12)    & 0.0003(21)  \\ 
        D96    & 0.745(14)    & -0.1971(45)    & -0.0007(10)    & -0.0018(22)  \\
        E112   & 0.747(12)    & -0.1980(45)    & -0.00150(95)    & -0.0032(21)  \\ \hline
        $a= 0$ & 0.753(18)(5) & -0.1973(61)(3) & -0.0010(15)(4)  & -0.0050(37)(31)  \\ \hline \hline
		\end{tabular*}
	}
	\vspace*{-0.2cm}
\end{table*}

\section{Comparison with other results}

In Tables~\ref{tbl:gA} and~\ref{tbl:gT}, we give our results for each gauge ensemble as well as our final continuum limit extrapolated results for the axial and tensor charges, respectively. The continuum values are given with statistical and systematic errors.

\begin{figure}[htbp]
    \centering
    \begin{minipage}{\columnwidth}
        \centering
        \includegraphics[width=0.8\textwidth]{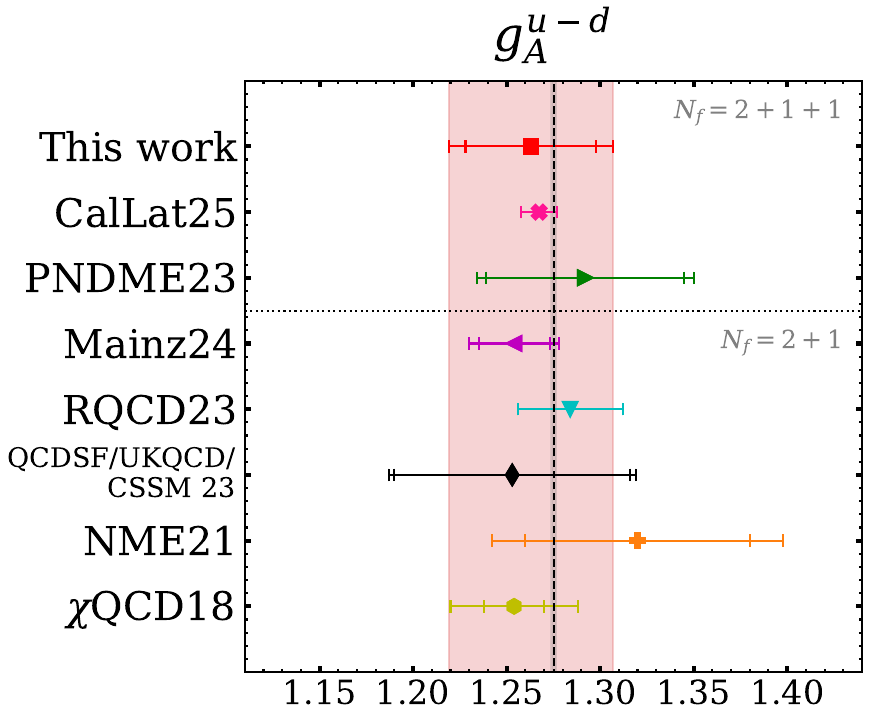}
        \vspace{0.5em}
        \includegraphics[width=0.8\textwidth]{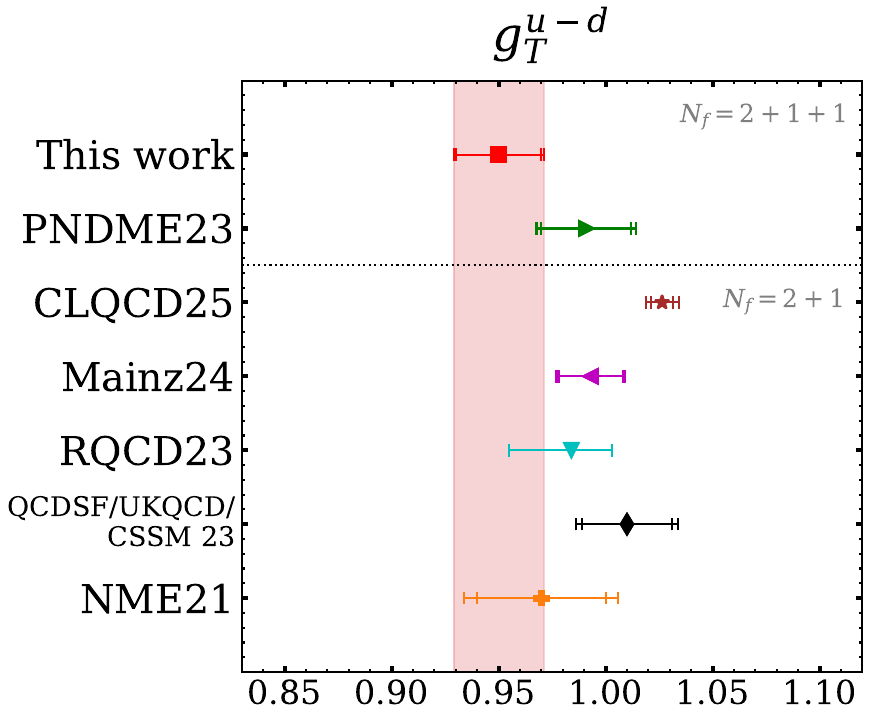}
        \caption{Comparison of the results of this work with other lattice QCD results, for the isovector axial (top) and tensor (bottom) charges. Our results are shown with the red square and red error band. The pink cross shows the result from CalLat~\cite{Hall:2025ytt},  the magenta triangles from the Mainz group~\cite{Djukanovic:2024krw}, the green triangles from PNDME~\cite{Jang:2023zts}, the brown star from CLQCD~\cite{Wang:2025nsd}, the cyan triangles from RQCD~\cite{Bali:2023sdi}, the black diamonds from the QCDSF/UKQCD/CSSM collaboration~\cite{QCDSFUKQCDCSSM:2023qlx}, the orange crosses from NME~\cite{Park:2021ypf} and the yellow hexagon from $\chi$QCD~\cite{Liang:2018pis}. For $\gaumd$, the dashed line and gray band represents the PDG value~\cite{ParticleDataGroup:2026mpi}. }
        \label{fig:umd_comb}
    \end{minipage}
\end{figure}

\begin{figure}[htbp]
	\includegraphics[width=\columnwidth]{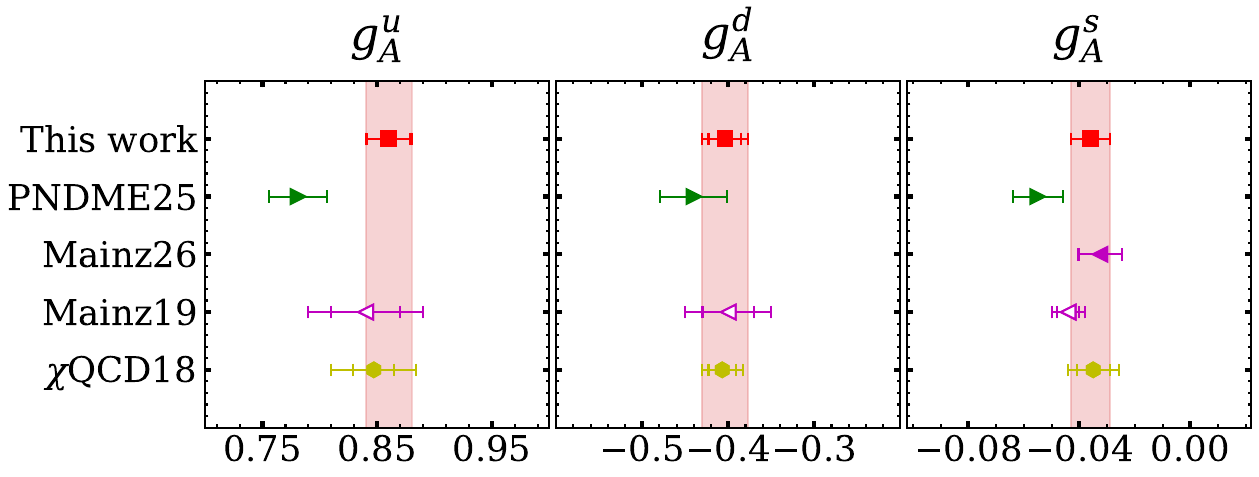}\\
	\vspace*{-0.3cm}
	\caption{Comparison of our results for the up- (left), down- (center), and strange-quark (right) contributions to the  axial charge with other lattice QCD  studies. Our results are shown with the red square and red error band. The green triangles show results from PNDME~\cite{Park:2025rxi} and the yellow hexagons from $\chi$QCD~\cite{Liang:2018pis}. For the Mainz group the magenta filled triangles show results from~\cite{Barone:2026uyx}, while the magenta open triangles from conference proceedings~\cite{Djukanovic:2019gvi}.}
	\vspace*{-0.4cm}
        \label{fig:gA_flavs_comb}
\end{figure}

\begin{figure}[htbp]
	\includegraphics[width=\columnwidth]{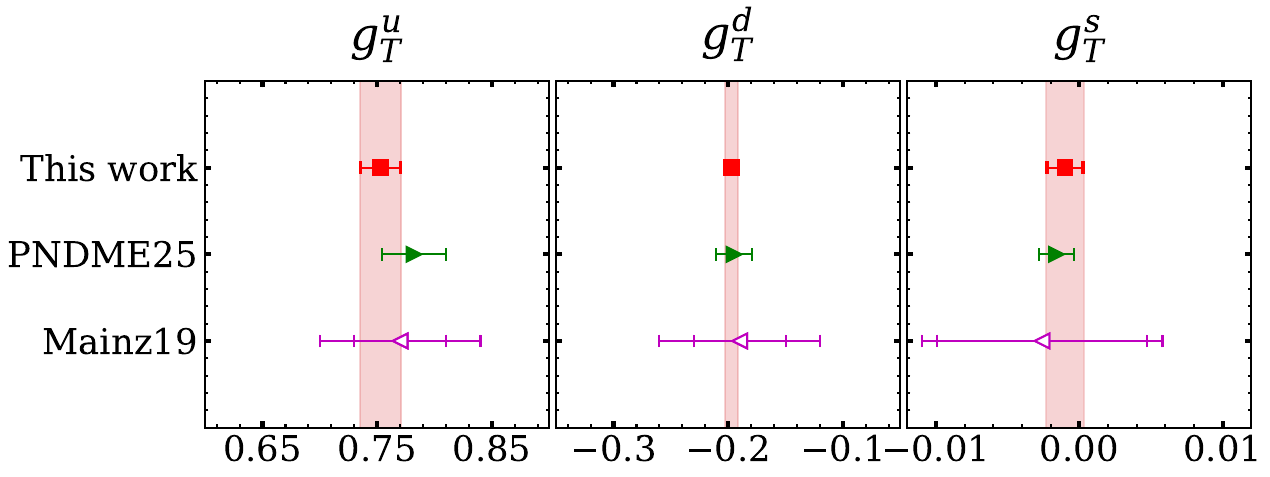}\\
	\vspace*{-0.3cm}
	\caption{Comparison of our results for the up- (left), down- (center), and strange-quark (right) contributions to the tensor charges with other lattice QCD analyses. Our results are shown with the red square and red error band. The green triangles show results from PNDME~\cite{Park:2025rxi} and the magenta open triangles from the Mainz group from conference proceedings~\cite{Djukanovic:2019gvi}.}
	\vspace*{-0.4cm}
        \label{fig:gT_flavs_comb}
\end{figure}

Our continuum extrapolated results  are compared  with other recent lattice QCD calculations. We emphasize that our results are the only ones among those considered here for which the continuum limit is obtained exclusively from ensembles simulated directly at the physical pion mass. In contrast, the other studies rely either entirely on ensembles with heavier-than-physical pion masses or on a combination of one or two ensembles simulated at physical pion mass with ensembles at heavier pion masses. Since  simulations at physical pion mass generally have larger statistical uncertainties, the corresponding continuum and chiral extrapolations can be more strongly constrained by the heavier-than-physical ensembles than by the data using the ensembles with physical pion mass. Given that chiral extrapolations of nucleon observables are known to be less reliable than those in the meson sector~\cite{Shanahan:2016pla}, such an approach can introduce additional systematic uncertainties that are difficult to quantify. Our analysis avoids this source of uncertainty by using ensembles simulated directly at the physical pion mass and therefore does not require a chiral extrapolation.

In Figs. \ref{fig:umd_comb}, \ref{fig:gA_flavs} and \ref{fig:gT_flavs} we compare our results for the nucleon charges  with values from other collaborations. 
Below we give some details on these recent lattice QCD studies.

\begin{itemize}

    \item The CalLat collaboration determined $\gaumd$ using an approach based on the Feynman-Hellmann theorem~\cite{Hall:2025ytt}. They used a mixed-action setup with $N_f{=}2{+}1{+}1$ highly improved staggered quarks (HISQ) in the sea, generated by the MILC collaboration, and M\"obius domain-wall fermions as valence quarks, with the gauge links smoothed using gradient flow. The analysis includes eighteen ensembles spanning three lattice spacings ($a \approx 0.09,~0.12,~0.15$ fm) and pion masses in the range of approximately $130$--$400$ MeV. The nucleon axial charge is extracted from multi-state fits to the nucleon two-point and Feynman-Hellmann correlation functions. The final value, denoted as CalLat25 in Fig.~\ref{fig:umd_comb}, was obtained through a model-averaged analysis accounting for chiral,  discretization, and finite-volume effects.

    \item The PNDME collaboration analyzed thirteen $N_f {=} 2{+}1{+}1$ ensembles, which belong to the same set used by CalLat, using highly improved staggered quarks (HISQ) in the sea and clover-improved valence quarks in a mixed-action setup for the computation of the nucleon correlation functions. The ensembles cover four lattice spacings, 0.06~fm, 0.09~fm, 0.12~fm and 0.15~fm, three pion masses (135~MeV, 220~MeV and 310~MeV), and volumes with $3.7 \leq m_\pi L \leq 5.5$. Two of these ensembles are at the physical pion mass. A combined chiral, continuum and finite-volume fit was performed to obtain the isovector nucleon charges at the physical point. Their results are denoted as PNDME23~\cite{Jang:2023zts} in Fig.~\ref{fig:umd_comb}.
    
    The PNDME collaboration has recently reported updated results for the flavor-diagonal up, down and strange axial and tensor charges~\cite{Park:2025rxi}. The calculation includes both connected and disconnected contributions and incorporates an analysis of excited-state effects including low-lying $N\pi$ states. The final results are obtained from a simultaneous chiral, continuum and finite-volume analysis. These results are denoted as PNDME25 in Figs.~\ref{fig:gA_flavs_comb} and~\ref{fig:gT_flavs_comb}.

    \item The RQCD collaboration performed an extensive study of nucleon charges~\cite{Bali:2023sdi}, analyzing forty-seven $N_f {=} 2{+}1$ ensembles with clover-improved fermions, most of which were generated within the Coordinated Lattice Simulations (CLS) effort. The ensembles cover pion masses from 130~MeV to 430~MeV, six lattice spacings ranging from approximately 0.039~fm to 0.098~fm, and volumes with $3.0 \leq m_\pi L \leq 6.5$, including two ensembles with pion masses close to the physical value. Extrapolations to the physical point were performed using leading-order chiral expressions for the pion-mass dependence, the leading asymptotic form for finite-volume corrections, and terms up to $a^2$ in the lattice spacing. Their final values are denoted by RQCD23 in Fig.~\ref{fig:umd_comb}.

    \item The Mainz collaboration investigated the isovector charges in Ref.~\cite{Djukanovic:2024krw} using fifteen $N_f {=} 2{+}1$ CLS ensembles, covering pion masses from 130~MeV to 360~MeV, four lattice spacings, and volumes with $3.05 \leq m_\pi L \leq 5.89$. A simultaneous extrapolation in the lattice spacing, pion mass, and volume was performed, with one ensemble having a pion mass slightly below the physical value. Their isovector charge results are denoted by Mainz24 in Fig.~\ref{fig:umd_comb}.
    
    In addition, the Mainz collaboration reported a determination of the up, down, and strange axial and tensor charges in conference proceedings~\cite{Djukanovic:2019gvi} on a subset of eight ensembles. These ensembles have four lattice spacings between 0.050~fm and 0.086~fm, volumes with $3.78 \leq m_\pi L \leq 5.29$, and pion masses in the range 200~MeV--360~MeV. No  ensembles with physical pion mass were included in this analysis. A combined chiral and continuum-limit fit was used to obtain the physical-point results. These results are denoted by Mainz19 in Figs.~\ref{fig:gA_flavs_comb} and~\ref{fig:gT_flavs_comb}. More recently, the Mainz collaboration published a determination of the strange axial form factor~\cite{Barone:2026uyx}. This calculation uses a subset of fourteen ensembles with pion masses in the range 130~MeV--350~MeV and four lattice spacings ranging from 0.050~fm to 0.086~fm. The strange axial charge is obtained from the strange axial form factor in the $Q^2\to0$ limit. In Fig.~\ref{fig:gA_flavs_comb}, we denote this newer strange axial result as Mainz26.

    \item The QCDSF/UKQCD/CSSM collaboration~\cite{QCDSFUKQCDCSSM:2023qlx} analyzed twenty-one $N_f {=} 2{+}1$ gauge ensembles, employing a tree-level Symanzik improved gluon action together with nonperturbatively $\mathcal{O}(a)$ clover-improved Wilson fermions. The ensembles cover pion masses in the range 220~MeV $\leq m_\pi \leq$ 468~MeV, five lattice spacings, and three lattice volumes. Combined extrapolations in the pion mass, lattice spacing, and volume were used to obtain their isovector charges, shown in Fig.~\ref{fig:umd_comb}. 
    
    \item The NME collaboration determined the isovector charges in Ref.~\cite{Park:2021ypf} using seven $N_f {=} 2{+}1$ ensembles with clover-improved Wilson fermions. The ensembles span pion masses of 166~MeV $\leq m_\pi \leq$ 285~MeV, lattice spacings from 0.071~fm to 0.127~fm, and volumes satisfying $3.75 \leq m_\pi L \leq 6.15$. The physical-point results were obtained through combined chiral, continuum, and infinite-volume extrapolations. These results are denoted by NME21 in Fig.~\ref{fig:umd_comb}.

    \item The CLQCD collaboration, recently, reported a high-precision determination of the isovector tensor charge~\cite{Wang:2025nsd}. They analyzed fifteen $N_f{=}2{+}1$ lattice ensembles, employing tadpole-improved tree-level Symanzik and Clover actions for the gauge and fermion sectors, respectively. The ensembles span five lattice spacings, $0.052~\mathrm{fm}\leq a\leq0.105~\mathrm{fm}$, with pion masses in the range $134~\mathrm{MeV}\leq m_\pi\leq341~\mathrm{MeV}$ and volumes with $2.67\leq m_\pi L\leq5.79$. Five sets of ensembles have the same quark masses and lattice spacing but different volumes, including three ensembles at the physical pion mass. Their final result includes systematic uncertainties from the continuum, infinite-volume and chiral extrapolations, as well as excited-state contamination and renormalization, and are denoted as CLQCD25 in Fig.~\ref{fig:umd_comb}.

    \item The $\chi$QCD collaboration determined the isovector as well as the up, down, and strange axial charges~\cite{Liang:2018pis} using a hybrid setup with domain wall fermions in the sea and overlap fermions in the valence sector. Their analysis included three RBC/UKQCD $N_f {=} 2 {+} 1$ gauge ensembles at three lattice spacings, $a=0.08$~fm, $a=0.11$~fm, and $a=0.14$~fm, two volumes, and three pion masses of 171~MeV, 302~MeV, and 337~MeV. Simultaneous continuum, chiral, and volume extrapolations were performed, with the resulting values denoted by $\chi$QCD18 in Figs.~\ref{fig:umd_comb} and \ref{fig:gA_flavs_comb}.

\end{itemize}

In Fig.~\ref{fig:umd_comb}, we compare our results for the isovector axial and tensor charges with recent lattice QCD calculations. For the isovector axial charge, our result is in good agreement with the results from the other collaborations and  compatible with the experimental determination~\cite{ParticleDataGroup:2026mpi}. The isovector tensor charge shows a somewhat larger spread among the lattice calculations. Our result agrees with all other  results within one standard deviation, except with the CLQCD value, which is larger than ours.

For the flavor-diagonal charges, shown in Figs.~\ref{fig:gA_flavs_comb} and \ref{fig:gT_flavs_comb}, we find overall good agreement with the available lattice results. The tensor charges show particularly good consistency among the different calculations. For the axial charges, our results are also generally consistent with the other determinations. The main visible difference is in the up-quark axial charge, where the PNDME result is smaller than ours.  The PNDME collaboration obtains their quoted up and down axial charges by averaging the results of their ``standard'' and ``$N\pi$'' analyses, a procedure not followed by the other collaborations. 

\begin{figure}[htbp]
    \raggedright
    \includegraphics[width=0.9\columnwidth]{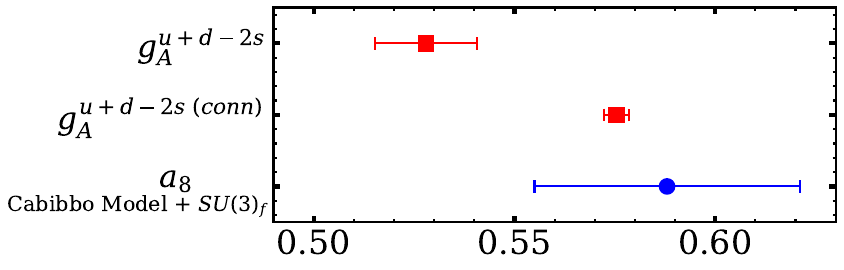}
    \vspace*{-0.3cm}
    \caption{Comparison of our result for the non-singlet axial charge combination $g_A^{u+d-2s}$ with the value obtained only from the connected contributions, and the value of $a_8$ extracted from hyperon decays using the Cabibbo Model and assuming SU(3)-flavor (SU(3)$_{\!f})$ symmetry~\cite{Close:1993mv,Deur:2018roz}. The red squares show our results for $g_A^{u+d-2s}$, while the blue circle shows the phenomenological value of $a_8$.}
    \label{fig:gA_8_comb}
\end{figure}

In Fig.~\ref{fig:gA_8_comb}, we present our result for the non-singlet axial charge combination $g_A^{u+d-2s}$, together with the corresponding value obtained exclusively from the connected contributions. The latter allows for a  comparison with the value of $a_8$ extracted from hyperon decays~\cite{Close:1993mv,Deur:2018roz}, using the Cabibbo Model and under the assumption of SU(3)-flavor symmetry. In the SU(3)$_{\!f}$-symmetric limit, the disconnected contributions cancel in this combination, motivating the comparison with the result obtained from the connected only. This provides a useful consistency check between our lattice determination and the phenomenological extraction from hyperon decays, while showing the effects of SU(3)-flavor symmetry breaking. 

\section{Summary and Conclusions}
\label{sec:conclusions}

The nucleon axial and tensor charges are determined using four $N_f=2+1+1$ twisted-mass clover-improved ensembles with quark masses tuned closely to their physical values. The calculation includes both connected and disconnected contributions, allowing us to determine the isovector, isoscalar, and individual flavor contributions to the axial and tensor charges. These results provide an update to our previous analysis~\cite{Alexandrou:2024ozj} using three ensembles, yielding smaller errors and more stability to the continuum extrapolation. 

Our results show  overall good agreement with other recent lattice QCD determinations. In particular, the isovector axial charge is consistent with both the available lattice results and the experimental determination. The flavor-diagonal axial and tensor charges are also generally consistent with the results from other lattice calculations, while the isovector tensor charge shows a somewhat larger spread among the different determinations.

The continuum limit is obtained using a model average of constant and linear-in-$a^2$ fits. Although the lattice-spacing dependence is mild, the extrapolation to $a=0$ has a non-negligible effect and is therefore important for obtaining the physical values, together with a comprehensive treatment of the systematic uncertainties due to excited-state contamination. Importantly, the continuum extrapolation is performed using ensembles simulated at approximately the physical pion mass, avoiding the need for a chiral extrapolation and the possible associated uncontrolled systematics.

\section*{ACKNOWLEDGMENTS}
We thank all members of the ETM Collaboration for the most enjoyable collaboration. 
We thank Roberto Frezzotti, Giuseppe Gagliardi and Vittorio Lubicz, for useful discussions and crosschecks on the analysis of the renormalization factors.
C.A., S.B, A.E, C.I., F.P. and G.S. acknowledge partial support from the projects IMAGE-N (EXCELLENCE/0524/0459), MuonHVP (EXCELLENCE/0524/0017), PulseQCD (EXCELLENCE/0524/0269), 
DeNuTra (EXCELLENCE/524/0455), HiQO (POST-DOC/0925-ROandOther/0148), Baryon8 (POSTDOC/0524/0001), $\phi$BSM (CULTURE/AWARD-YR/0524 B/0002), StrongILA (EXCELLENCE/0524/0001), and partonWF (VISION ERC/0525/0010),  funded by the European Regional Development Fund and the Republic of Cyprus through the Cyprus Research and Innovation Foundation.
C.A. acknowledges partial support from the European Joint Doctorate AQTIVATE that received funding from the European Union's research and innovation program under the Marie Sklodowska-Curie Doctoral Networks action, Grant Agreement No 101072344.
The authors gratefully acknowledge the Gauss Centre for Supercomputing e.V. (www.gauss-centre.eu) for funding this project by providing computing time through the John von Neumann Institute for Computing (NIC) on the GCS Supercomputers JUWELS, JUWELS Booster~\cite{JUWELS-BOOSTER} and JUPITER Booster at J\"ulich Supercomputing Centre (JSC).
The authors acknowledge access to the JUPITER supercomputer, which is funded by the EuroHPC Joint Undertaking, the German Federal Ministry of Research, Technology and Space, and the Ministry of Culture and Science of the German state of North Rhine-Westphalia, through the JUPITER Research and Early Access Program (JUREAP) as part of the EuroHPC project application EHPC-EXT-2023E02-052.

\newpage

\bibliography{refs}

@article{Detmold:2001jb,
    author = "Detmold, William and Melnitchouk, W. and Negele, John W. and Renner, Dru Bryant and Thomas, Anthony William",
    title = "{Chiral extrapolation of lattice moments of proton quark distributions}",
    eprint = "hep-lat/0103006",
    archivePrefix = "arXiv",
    reportNumber = "ADP-01-07-T442, JLAB-THY-01-09, MIT-CTP-3092, JLAB-THY-01-43",
    doi = "10.1103/PhysRevLett.87.172001",
    journal = "Phys. Rev. Lett.",
    volume = "87",
    pages = "172001",
    year = "2001"
}

@article{Cocuzza:2023oam,
    author = "Cocuzza, C. and Metz, A. and Pitonyak, D. and Prokudin, A. and Sato, N. and Seidl, R.",
    collaboration = "JAM",
    title = "{Transversity Distributions and Tensor Charges of the Nucleon: Extraction from Dihadron Production and Their Universal Nature}",
    eprint = "2306.12998",
    archivePrefix = "arXiv",
    primaryClass = "hep-ph",
    reportNumber = "JLAB-THY-23-3859",
    doi = "10.1103/PhysRevLett.132.091901",
    journal = "Phys. Rev. Lett.",
    volume = "132",
    number = "9",
    pages = "091901",
    year = "2024"
}

@article{Undagoitia:2015gya,
author = "Marrod\'an Undagoitia, Teresa and Rauch, Ludwig",
archivePrefix = "arXiv",
doi = "10.1088/0954-3899/43/1/013001",
eprint = "1509.08767",
journal = "J. Phys. G",
number = "1",
pages = "013001",
primaryClass = "physics.ins-det",
title = "{Dark matter direct-detection experiments}",
volume = "43",
year = "2016"
}

@article{Chizhov:1995wp,
    author = "Chizhov, M. V.",
    title = "{Search for tensor interactions in kaon decays at DAPHNE}",
    eprint = "hep-ph/9511287",
    archivePrefix = "arXiv",
    reportNumber = "JINR-E2-95-457",
    doi = "10.1016/0370-2693(96)00535-7",
    journal = "Phys. Lett. B",
    volume = "381",
    pages = "359--364",
    year = "1996"
}

@article{Dhargyal:2016jgo,
	author = "Dhargyal, Lobsang",
	editor = "Naimuddin, Md.",
	archivePrefix = "arXiv",
	doi = "10.1007/978-3-319-73171-1\_75",
	eprint = "1610.06293",
	journal = "Springer Proc. Phys.",
	pages = "329--331",
	primaryClass = "hep-ph",
	title = "{New tensor interaction as the source of the observed CP asymmetry in $\tau \to K_{S}\pi\nu_{\tau}$}",
	volume = "203",
	year = "2018"
}

@article{Bischer:2018zcz,
    author = "Bischer, Ingolf and Rodejohann, Werner",
    title = "{General Neutrino Interactions at the DUNE Near Detector}",
    eprint = "1810.02220",
    archivePrefix = "arXiv",
    primaryClass = "hep-ph",
    doi = "10.1103/PhysRevD.99.036006",
    journal = "Phys. Rev. D",
    volume = "99",
    number = "3",
    pages = "036006",
    year = "2019"
}

@article{COHERENT:2017ipa,
    author = "Akimov, D. and others",
    collaboration = "COHERENT",
    title = "{Observation of Coherent Elastic Neutrino-Nucleus Scattering}",
    eprint = "1708.01294",
    archivePrefix = "arXiv",
    primaryClass = "nucl-ex",
    doi = "10.1126/science.aao0990",
    journal = "Science",
    volume = "357",
    number = "6356",
    pages = "1123--1126",
    year = "2017"
}

@article{Beda:2012zz,
	author = "Beda, A.G. and Brudanin, V.B. and Egorov, V.G. and Medvedev, D.V. and Pogosov, V.S. and Shirchenko, M.V. and Starostin, A.S.",
	doi = "10.1155/2012/350150",
	journal = "Adv. High Energy Phys.",
	pages = "350150",
	title = "{The results of search for the neutrino magnetic moment in GEMMA experiment}",
	volume = "2012",
	year = "2012"
}

@article{Gockeler:1998ye,
	author = "Gockeler, M. and Horsley, R. and Oelrich, H. and Perlt, H. and Petters, D. and Rakow, Paul E.L. and Schafer, A. and Schierholz, G. and Schiller, A.",
	archivePrefix = "arXiv",
	doi = "10.1016/S0550-3213(99)00036-X",
	eprint = "hep-lat/9807044",
	journal = "Nucl. Phys. B",
	pages = "699--733",
	reportNumber = "DESY-98-097, TPR-98-19, HUB-EP-98-45",
	title = "{Nonperturbative renormalization of composite operators in lattice QCD}",
	volume = "544",
	year = "1999"
}

@article{Bar:2018xyi,
    author = {B\"ar, Oliver},
    title = "{$N\pi$-state contamination in lattice calculations of the nucleon axial form factors}",
    eprint = "1812.09191",
    archivePrefix = "arXiv",
    primaryClass = "hep-lat",
    reportNumber = "HU-EP-18/40",
    doi = "10.1103/PhysRevD.99.054506",
    journal = "Phys. Rev. D",
    volume = "99",
    number = "5",
    pages = "054506",
    year = "2019"
}

@article{Bhattacharya:2011qm,
	author         = "Bhattacharya, Tanmoy and Cirigliano, Vincenzo and Cohen,
	Saul D. and Filipuzzi, Alberto and Gonzalez-Alonso, Martin
	and Graesser, Michael L. and Gupta, Rajan and Lin,
	Huey-Wen",
	title          = "{Probing Novel Scalar and Tensor Interactions from
	(Ultra)Cold Neutrons to the LHC}",
	journal        = "Phys. Rev.",
	volume         = "D85",
	year           = "2012",
	pages          = "054512",
	doi            = "10.1103/PhysRevD.85.054512",
	eprint         = "1110.6448",
	archivePrefix  = "arXiv",
	primaryClass   = "hep-ph",
	reportNumber   = "LA-UR-11-11460, NT@UW-11-16, IFIC-11-57, FTUV-11-1007,
	NPAC-11-14",
	SLACcitation   = "%%CITATION = ARXIV:1110.6448;%%"
}

@article{Frezzotti:2000nk,
	author         = "Frezzotti, Roberto and Grassi, Pietro Antonio and Sint,
	Stefan and Weisz, Peter",
	title          = "{Lattice QCD with a chirally twisted mass term}",
	collaboration  = "Alpha",
	journal        = "JHEP",
	volume         = "08",
	year           = "2001",
	pages          = "058",
	eprint         = "hep-lat/0101001",
	archivePrefix  = "arXiv",
	primaryClass   = "hep-lat",
	reportNumber   = "CERN-TH-2000-384, MPI-PHT-2000-51, BICOCCA-FT-0027,
	NYU-TH-00-09-13",
	SLACcitation   = "%%CITATION = HEP-LAT/0101001;%%"
}

@article{Sheikholeslami:1985ij,
	author         = "Sheikholeslami, B. and Wohlert, R.",
	title          = "{Improved Continuum Limit Lattice Action for QCD with
	Wilson Fermions}",
	journal        = "Nucl. Phys.",
	volume         = "B259",
	year           = "1985",
	pages          = "572",
	doi            = "10.1016/0550-3213(85)90002-1",
	reportNumber   = "DESY-85-024",
	SLACcitation   = "%%CITATION = NUPHA,B259,572;%%"
}

@article{Alexandrou:2018sjm,
    author = "Alexandrou, C. and Bacchio, S. and Constantinou, M. and Finkenrath, J. and Hadjiyiannakou, K. and Jansen, K. and Koutsou, G. and Vaquero Aviles-Casco, A.",
    title = "{Proton and neutron electromagnetic form factors from lattice QCD}",
    eprint = "1812.10311",
    archivePrefix = "arXiv",
    primaryClass = "hep-lat",
    reportNumber = "DESY-18-033, DESY 18-033",
    doi = "10.1103/PhysRevD.100.014509",
    journal = "Phys. Rev. D",
    volume = "100",
    number = "1",
    pages = "014509",
    year = "2019"
}

@article{Alexandrou:2018egz,
	author         = "Alexandrou, Constantia and others",
	title          = "{Simulating twisted mass fermions at physical light,
	strange and charm quark masses}",
	journal        = "Phys. Rev.",
	volume         = "D98",
	year           = "2018",
	number         = "5",
	pages          = "054518",
	doi            = "10.1103/PhysRevD.98.054518",
	eprint         = "1807.00495",
	archivePrefix  = "arXiv",
	primaryClass   = "hep-lat",
	SLACcitation   = "%%CITATION = ARXIV:1807.00495;%%"
}

@article{McNeile:2006bz,
      author         = "McNeile, C. and Michael, Christopher",
      title          = "{Decay width of light quark hybrid meson from the
                        lattice}",
      collaboration  = "UKQCD",
      journal        = "Phys. Rev.",
      volume         = "D73",
      year           = "2006",
      pages          = "074506",
      doi            = "10.1103/PhysRevD.73.074506",
      eprint         = "hep-lat/0603007",
      archivePrefix  = "arXiv",
      primaryClass   = "hep-lat",
      reportNumber   = "LTH-691",
      SLACcitation   = "%%CITATION = HEP-LAT/0603007;%%"
}

@article{Stathopoulos:2013aci,
    author = "Stathopoulos, Andreas and Laeuchli, Jesse and Orginos, Kostas",
    title = "{Hierarchical Probing for Estimating the Trace of the Matrix Inverse on Toroidal Lattices}",
    eprint = "1302.4018",
    archivePrefix = "arXiv",
    primaryClass = "hep-lat",
    reportNumber = "JLAB-THY-13-1725",
    doi = "10.1137/120881452",
    journal = "SIAM J. Sci. Comput.",
    volume = "35",
    number = "5",
    pages = "S299--S322",
    year = "2013"
}

@article{Gusken:1989qx,
    author = "Gusken, S.",
    editor = "Cabibbo, N. and Marinari, E. and Parisi, G. and Petronzio, Roberto and Maiani, L. and Martinelli, G. and Pettorino, Roberto",
    title = "{A Study of smearing techniques for hadron correlation functions}",
    doi = "10.1016/0920-5632(90)90273-W",
    journal = "Nucl. Phys. B Proc. Suppl.",
    volume = "17",
    pages = "361--364",
    year = "1990"
}

@article{Alexandrou:1992ti,
      author         = "Alexandrou, C. and Gusken, S. and Jegerlehner, F. and
                        Schilling, K. and Sommer, R.",
      title          = "{The Static approximation of heavy - light quark systems:
                        A Systematic lattice study}",
      journal        = "Nucl. Phys.",
      volume         = "B414",
      year           = "1994",
      pages          = "815-855",
      doi            = "10.1016/0550-3213(94)90262-3",
      eprint         = "hep-lat/9211042",
      archivePrefix  = "arXiv",
      primaryClass   = "hep-lat",
      reportNumber   = "PSI-PR-92-27, CERN-TH-6692-92, WUB-92-34",
      SLACcitation   = "%%CITATION = HEP-LAT/9211042;%%"
}

@article{Albanese:1987ds,
      author         = "Albanese, M. and others",
      title          = "{Glueball Masses and String Tension in Lattice QCD}",
      collaboration  = "APE",
      journal        = "Phys. Lett.",
      volume         = "B192",
      year           = "1987",
      pages          = "163-169",
      doi            = "10.1016/0370-2693(87)91160-9",
      reportNumber   = "ROM2F-87-014",
      SLACcitation   = "%%CITATION = PHLTA,B192,163;%%"
}

@article{Gambhir:2016uwp,
      author         = "Gambhir, Arjun Singh and Stathopoulos, Andreas and
                        Orginos, Kostas",
      title          = "{Deflation as a Method of Variance Reduction for
                        Estimating the Trace of a Matrix Inverse}",
      journal        = "SIAM J. Sci. Comput.",
      volume         = "39",
      year           = "2017",
      pages          = "A532-A558",
      doi            = "10.1137/16M1066361",
      eprint         = "1603.05988",
      archivePrefix  = "arXiv",
      primaryClass   = "hep-lat",
      reportNumber   = "JLAB-THY-16-2291",
      SLACcitation   = "%%CITATION = ARXIV:1603.05988;%%"
}

@article{Frezzotti:2003ni,
      author         = "Frezzotti, R. and Rossi, G. C.",
            title          = "{Chirally improving Wilson fermions. 1. O(a)
                        improvement}",
			      journal        = "JHEP",
			            volume         = "08",
				          year           = "2004",
					        pages          = "007",
						      doi            = "10.1088/1126-6708/2004/08/007",
						            eprint         = "hep-lat/0306014",
							          archivePrefix  = "arXiv",
								        primaryClass   = "hep-lat",
									      reportNumber   = "BICOCCA-FT-03-15, ROM2F-2003-15",
									            SLACcitation   = "%%CITATION = HEP-LAT/0306014;%%"
										    }

@article{Alexandrou:2015sea,
      author         = "Alexandrou, Constantia and Constantinou, Martha and
                        Panagopoulos, Haralambos",
      title          = "{Renormalization functions for Nf=2 and Nf=4 twisted mass
                        fermions}",
      collaboration  = "ETM",
      journal        = "Phys. Rev.",
      volume         = "D95",
      year           = "2017",
      number         = "3",
      pages          = "034505",
      doi            = "10.1103/PhysRevD.95.034505",
      eprint         = "1509.00213",
      archivePrefix  = "arXiv",
      primaryClass   = "hep-lat",
      SLACcitation   = "%%CITATION = ARXIV:1509.00213;%%"
}

@article{Martinelli:1994ty,
	author         = "Martinelli, G. and Pittori, C. and Sachrajda, Christopher
	T. and Testa, M. and Vladikas, A.",
	title          = "{A General method for nonperturbative renormalization of
	lattice operators}",
	journal        = "Nucl. Phys.",
	volume         = "B445",
	year           = "1995",
	pages          = "81-108",
	doi            = "10.1016/0550-3213(95)00126-D",
	eprint         = "hep-lat/9411010",
	archivePrefix  = "arXiv",
	primaryClass   = "hep-lat",
	reportNumber   = "CERN-TH-7342-94, LPTHE-ORSAY-94-52, ROME-1022-1994,
	SHEP-94-95-03",
	SLACcitation   = "%%CITATION = HEP-LAT/9411010;%%"
}

@article{Constantinou:2016ieh,
	author         = "Constantinou, M. and Hadjiantonis, M. and Panagopoulos,
	H. and Spanoudes, G.",
	title          = "{Singlet versus nonsinglet perturbative renormalization
	of fermion bilinears}",
	journal        = "Phys. Rev.",
	volume         = "D94",
	year           = "2016",
	number         = "11",
	pages          = "114513",
	doi            = "10.1103/PhysRevD.94.114513",
	eprint         = "1610.06744",
	archivePrefix  = "arXiv",
	primaryClass   = "hep-lat",
	SLACcitation   = "%%CITATION = ARXIV:1610.06744;%%"
}

@article{Alexandrou:2019ali,
    author = "Alexandrou, C. and others",
    title = "{Moments of nucleon generalized parton distributions from lattice QCD simulations at physical pion mass}",
    eprint = "1908.10706",
    archivePrefix = "arXiv",
    primaryClass = "hep-lat",
    doi = "10.1103/PhysRevD.101.034519",
    journal = "Phys. Rev. D",
    volume = "101",
    number = "3",
    pages = "034519",
    year = "2020"
}

@article{Finkenrath:2022eon,
    author = "Finkenrath, Jacob and others",
    title = "{Twisted mass gauge ensembles at physical values of the light, strange and charm quark masses}",
    eprint = "2201.02551",
    archivePrefix = "arXiv",
    primaryClass = "hep-lat",
    doi = "10.22323/1.396.0284",
    journal = "PoS",
    volume = "LATTICE2021",
    pages = "284",
    year = "2022"
}

@article{ExtendedTwistedMass:2021gbo,
    author = "Alexandrou, C. and others",
    collaboration = "Extended Twisted Mass",
    title = "{Quark masses using twisted-mass fermion gauge ensembles}",
    eprint = "2104.13408",
    archivePrefix = "arXiv",
    primaryClass = "hep-lat",
    doi = "10.1103/PhysRevD.104.074515",
    journal = "Phys. Rev. D",
    volume = "104",
    number = "7",
    pages = "074515",
    year = "2021"
}

@article{Jay:2020jkz,
    author = "Jay, William I. and Neil, Ethan T.",
    title = "{Bayesian model averaging for analysis of lattice field theory results}",
    eprint = "2008.01069",
    archivePrefix = "arXiv",
    primaryClass = "stat.ME",
    reportNumber = "FERMILAB-PUB-20-374-T",
    doi = "10.1103/PhysRevD.103.114502",
    journal = "Phys. Rev. D",
    volume = "103",
    pages = "114502",
    year = "2021"
}

@article{Neil:2022joj,
    author = "Neil, Ethan T. and Sitison, Jacob W.",
    title = "{Improved information criteria for Bayesian model averaging in lattice field theory}",
    eprint = "2208.14983",
    archivePrefix = "arXiv",
    primaryClass = "stat.ME",
    doi = "10.1103/PhysRevD.109.014510",
    journal = "Phys. Rev. D",
    volume = "109",
    number = "1",
    pages = "014510",
    year = "2024"
}

@article{Jang:2023zts,
    author = "Jang, Yong-Chull and Gupta, Rajan and Bhattacharya, Tanmoy and Yoon, Boram and Lin, Huey-Wen",
    collaboration = "Precision Neutron Decay Matrix Elements (PNDME)",
    title = "{Nucleon isovector axial form factors}",
    eprint = "2305.11330",
    archivePrefix = "arXiv",
    primaryClass = "hep-lat",
    reportNumber = "Los Alamos LA-UR-23-25225",
    doi = "10.1103/PhysRevD.109.014503",
    journal = "Phys. Rev. D",
    volume = "109",
    number = "1",
    pages = "014503",
    year = "2024"
}

@article{Bali:2023sdi,
    author = {Bali, Gunnar S. and Collins, Sara and Heybrock, Simon and L\"offler, Marius and R\"odl, Rudolf and S\"oldner, Wolfgang and Weish\"aupl, Simon},
    collaboration = "RQCD",
    title = "{Octet baryon isovector charges from Nf=2+1 lattice QCD}",
    eprint = "2305.04717",
    archivePrefix = "arXiv",
    primaryClass = "hep-lat",
    doi = "10.1103/PhysRevD.108.034512",
    journal = "Phys. Rev. D",
    volume = "108",
    number = "3",
    pages = "034512",
    year = "2023"
}

@article{Djukanovic:2024krw,
    author = "Djukanovic, Dalibor and von Hippel, Georg and Meyer, Harvey B. and Ottnad, Konstantin and Wittig, Hartmut",
    title = "{Improved analysis of isovector nucleon matrix elements with Nf=2+1 flavors of O(a) improved Wilson fermions}",
    eprint = "2402.03024",
    archivePrefix = "arXiv",
    primaryClass = "hep-lat",
    reportNumber = "MITP-24-014",
    doi = "10.1103/PhysRevD.109.074507",
    journal = "Phys. Rev. D",
    volume = "109",
    number = "7",
    pages = "074507",
    year = "2024"
}

@article{QCDSFUKQCDCSSM:2023qlx,
    author = "Smail, R. E. and others",
    collaboration = "QCDSF/UKQCD/CSSM",
    title = "{Constraining beyond the standard model nucleon isovector charges}",
    eprint = "2304.02866",
    archivePrefix = "arXiv",
    primaryClass = "hep-lat",
    reportNumber = "ADP-23-10/T1219, DESY-23-047, LTH 1336, MIT-CTP/5581",
    doi = "10.1103/PhysRevD.108.094511",
    journal = "Phys. Rev. D",
    volume = "108",
    number = "9",
    pages = "094511",
    year = "2023"
}

@article{Djukanovic:2019gvi,
    author = "Djukanovic, Dalibor and Meyer, Harvey and Ottnad, Konstantin and von Hippel, Georg and Wilhelm, Jonas and Wittig, Hartmut",
    title = "{Strange nucleon form factors and isoscalar charges with $N_f=2+1$ $\mathcal{O}(a)$-improved Wilson fermions}",
    eprint = "1911.01177",
    archivePrefix = "arXiv",
    primaryClass = "hep-lat",
    reportNumber = "MITP/19-069",
    doi = "10.22323/1.363.0158",
    journal = "PoS",
    volume = "LATTICE2019",
    pages = "158",
    year = "2019"
}

@article{Bar:2016uoj,
    author = {B\"ar, Oliver},
    title = "{Nucleon-pion-state contribution in lattice calculations of the nucleon charges $g_A,g_T$ and $g_S$}",
    eprint = "1606.09385",
    archivePrefix = "arXiv",
    primaryClass = "hep-lat",
    reportNumber = "YITP-16-81",
    doi = "10.1103/PhysRevD.94.054505",
    journal = "Phys. Rev. D",
    volume = "94",
    number = "5",
    pages = "054505",
    year = "2016"
}

@article{Liang:2018pis,
    author = "Liang, Jian and Yang, Yi-Bo and Draper, Terrence and Gong, Ming and Liu, Keh-Fei",
    title = "{Quark spins and Anomalous Ward Identity}",
    eprint = "1806.08366",
    archivePrefix = "arXiv",
    primaryClass = "hep-ph",
    doi = "10.1103/PhysRevD.98.074505",
    journal = "Phys. Rev. D",
    volume = "98",
    number = "7",
    pages = "074505",
    year = "2018"
}

@article{Park:2021ypf,
    author = "Park, Sungwoo and Gupta, Rajan and Yoon, Boram and Mondal, Santanu and Bhattacharya, Tanmoy and Jang, Yong-Chull and Jo\'o, B\'alint and Winter, Frank",
    collaboration = "Nucleon Matrix Elements (NME)",
    title = "{Precision nucleon charges and form factors using (2+1)-flavor lattice QCD}",
    eprint = "2103.05599",
    archivePrefix = "arXiv",
    primaryClass = "hep-lat",
    reportNumber = "LA-UR-21-20526, JLAB-THY-22-3583",
    doi = "10.1103/PhysRevD.105.054505",
    journal = "Phys. Rev. D",
    volume = "105",
    number = "5",
    pages = "054505",
    year = "2022"
}

@article{JUWELS-BOOSTER,
author = {{J\"{u}lich Supercomputing Centre}},
title = {{JUWELS Cluster and Booster: Exascale Pathfinder with Modular Supercomputing Architecture at Juelich Supercomputing Centre}},
journal = {Journal of large-scale research facilities},
number = {A138},
volume = {7},
doi = {10.17815/jlsrf-7-183},
url = {http://dx.doi.org/10.17815/jlsrf-7-183},
year = {2021}
}

@article{Abdel-Rehim:2016pjw,
    author = "Abdel-Rehim, Abdou and Alexandrou, Constantia and Constantinou, Martha and Finkenrath, Jacob and Hadjiyiannakou, Kyriakos and Jansen, Karl and Kallidonis, Christos and Koutsou, Giannis and Avil\'es-Casco, Alejandro Vaquero and Volmer, Julia",
    title = "{Disconnected diagrams with twisted-mass fermions}",
    eprint = "1611.03802",
    archivePrefix = "arXiv",
    primaryClass = "hep-lat",
    doi = "10.22323/1.256.0155",
    journal = "PoS",
    volume = "LATTICE2016",
    pages = "155",
    year = "2016"
    }

@article{Shanahan:2016pla,
    author = "Shanahan, P. E.",
    title = "{Chiral Effective Theory Methods and their Application to the Structure of Hadrons from Lattice QCD}",
    eprint = "1606.08812",
    archivePrefix = "arXiv",
    primaryClass = "hep-lat",
    reportNumber = "MIT-CTP-4784",
    doi = "10.1088/0954-3899/43/12/124001",
    journal = "J. Phys. G",
    volume = "43",
    number = "12",
    pages = "124001",
    year = "2016"
}

@article{Chetyrkin:1997dh,
    author = "Chetyrkin, K. G.",
    title = "{Quark mass anomalous dimension to O (alpha-s**4)}",
    eprint = "hep-ph/9703278",
    archivePrefix = "arXiv",
    reportNumber = "MPI-PHT-97-019",
    doi = "10.1016/S0370-2693(97)00535-2",
    journal = "Phys. Lett. B",
    volume = "404",
    pages = "161--165",
    year = "1997"
}

@article{Gracey:2003yr,
    author = "Gracey, J. A.",
    title = "{Three loop anomalous dimension of nonsinglet quark currents in the RI-prime scheme}",
    eprint = "hep-ph/0304113",
    archivePrefix = "arXiv",
    reportNumber = "LTH-572",
    doi = "10.1016/S0550-3213(03)00335-3",
    journal = "Nucl. Phys. B",
    volume = "662",
    pages = "247--278",
    year = "2003"
}

@article{Gracey:2022vqr,
    author = "Gracey, J. A.",
    title = "{Tensor current renormalization in the RI' scheme at four loops}",
    eprint = "2208.14527",
    archivePrefix = "arXiv",
    primaryClass = "hep-ph",
    reportNumber = "LTH 1312",
    doi = "10.1103/PhysRevD.106.085008",
    journal = "Phys. Rev. D",
    volume = "106",
    number = "8",
    pages = "085008",
    year = "2022"
}

@inproceedings{Bernard:1985ss,
    author = "Bernard, Claude W. and Draper, Terrence and Hockney, G. and Soni, A.",
    title = "{CALCULATION OF WEAK MATRIX ELEMENTS: SOME TECHNICAL ASPECTS}",
    booktitle = "{Workshop on Lattice Gauge Theories}",
    year = "1985",
    month = "11",
    address = "Wuppertal, Germany"
}

@preprint{Alexandrou:2026oks,
    author = "Alexandrou, Constantia and Bacchio, Simone and Finkenrath, Jacob and Iona, Christos and Koutsou, Giannis and Kummer, Christian and Li, Yan and Prasad, Bhavna and Spanoudes, Gregoris",
    title = "{Nucleon unpolarized second Mellin moments using lattice QCD ensembles with physical quark masses and in the continuum limit}",
    eprint = "2607.20337",
    archivePrefix = "arXiv",
    primaryClass = "hep-lat",
    month = "7",
    year = "2026"
    }

@article{Alexandrou:2024ozj,
    author = "Alexandrou, C. and Bacchio, S. and Finkenrath, J. and Iona, C. and Koutsou, G. and Li, Y. and Spanoudes, G.",
    title = "{Nucleon charges and {\ensuremath{\sigma}}-terms in lattice QCD}",
    eprint = "2412.01535",
    archivePrefix = "arXiv",
    primaryClass = "hep-lat",
    doi = "10.1103/PhysRevD.111.054505",
    journal = "Phys. Rev. D",
    volume = "111",
    number = "5",
    pages = "054505",
    year = "2025"
}

@article{Alexandrou:2025vto,
    author = "Alexandrou, Constantia and Bacchio, Simone and Koutsou, Giannis and Prasad, Bhavna and Spanoudes, Gregoris",
    title = "{Proton and neutron electromagnetic form factors in the continuum limit using lattice QCD ensembles with physical pion masses}",
    eprint = "2507.20910",
    archivePrefix = "arXiv",
    primaryClass = "hep-lat",
    doi = "10.1103/tt39-n1df",
    journal = "Phys. Rev. D",
    volume = "113",
    number = "11",
    pages = "114524",
    year = "2026"
}

@preprint{Alexandrou:2026ait,
    author = "Alexandrou, Constantia and Bacchio, Simone and Bode, Mathis and Finkenrath, Jacob and Herten, Andreas and Iona, Christos and Koutsou, Giannis and Pittler, Ferenc and Prasad, Bhavna and Spanoudes, Gregoris",
    title = "{Nucleon strange electromagnetic form factors from $N_f=2+1+1$ lattice QCD}",
    eprint = "2603.26591",
    archivePrefix = "arXiv",
    primaryClass = "hep-lat",
    month = "3",
    year = "2026"
}

@article{ETMC-ren:2026,
    author = "{Extended Twisted Mass Collaboration}",
    title = "{Non-perturbative renormalisation of quark bilinear operators with  $N_f=2+1+1$ Wilson-clover twisted mass fermions}",
    journal = "In preparation"
}

@article{ExtendedTwistedMass:2024nyi,
    author = "Alexandrou, C. and others",
    collaboration = "Extended Twisted Mass",
    title = "{Strange and charm quark contributions to the muon anomalous magnetic moment in lattice QCD with twisted-mass fermions}",
    eprint = "2411.08852",
    archivePrefix = "arXiv",
    primaryClass = "hep-lat",
    reportNumber = "CERN-TH-2024-197",
    doi = "10.1103/PhysRevD.111.054502",
    journal = "Phys. Rev. D",
    volume = "111",
    number = "5",
    pages = "054502",
    year = "2025"
}

@article{Park:2025rxi,
    author = "Park, Sungwoo and Gupta, Rajan and Bhattacharya, Tanmoy and He, Fangcheng and Mondal, Santanu and Lin, Huey-Wen and Yoon, Boram",
    title = "{Flavor diagonal nucleon charges using clover fermions on MILC HISQ ensembles}",
    eprint = "2503.07100",
    archivePrefix = "arXiv",
    primaryClass = "hep-lat",
    reportNumber = "LLNL-JRNL-858665, LA-UR-22053",
    doi = "10.1103/hn5v-9zgm",
    journal = "Phys. Rev. D",
    volume = "112",
    number = "5",
    pages = "054508",
    year = "2025"
}

@article{Barone:2026uyx,
    author = "Barone, Alessandro and Djukanovic, Dalibor and von Hippel, Georg and Meyer, Harvey B. and Ottnad, Konstantin and Wittig, Hartmut",
    title = "{Strange and flavor-singlet axial form factors of the nucleon from lattice QCD}",
    eprint = "2605.06559",
    archivePrefix = "arXiv",
    primaryClass = "hep-lat",
    reportNumber = "MITP-26-022",
    doi = "10.1103/dgcm-4yz4",
    journal = "Phys. Rev. D",
    volume = "114",
    number = "3",
    pages = "034520",
    year = "2026"
}

@preprint{Wang:2025nsd,
    author = "Wang, Ji-Hao and Hu, Zhi-Cheng and Ji, Xiangdong and Jiang, Xiangyu and Su, Yushan and Sun, Peng and Yang, Yi-Bo",
    collaboration = "CLQCD",
    title = "{Precision determination of nucleon iso-vector scalar and tensor charges at the physical point}",
    eprint = "2511.02326",
    archivePrefix = "arXiv",
    primaryClass = "hep-lat",
    month = "11",
    year = "2025"
}

@article{ParticleDataGroup:2026mpi,
    author = "Takahashi, F. and others",
    collaboration = "Particle Data Group",
    title = "{Review of Particle Physics}",
    doi = "10.1142/s0217751x26300115",
    journal = "Int. J. Mod. Phys. A",
    volume = "41",
    number = "22",
    pages = "2630011",
    year = "2026"
}

@article{Hall:2025ytt,
    author = "Hall, Zack B. and others",
    title = "{Signs of nonmonotonic finite-volume corrections to gA}",
    eprint = "2503.09891",
    archivePrefix = "arXiv",
    primaryClass = "hep-lat",
    doi = "10.1103/q88h-2vz8",
    journal = "Phys. Rev. C",
    volume = "113",
    number = "5",
    pages = "055206",
    year = "2026"
}

@article{Close:1993mv,
    author = "Close, F. E. and Roberts, R. G.",
    title = "{Consistent analysis of the spin content of the nucleon}",
    eprint = "hep-ph/9306289",
    archivePrefix = "arXiv",
    reportNumber = "RAL-93-040",
    doi = "10.1016/0370-2693(93)90673-6",
    journal = "Phys. Lett. B",
    volume = "316",
    pages = "165--171",
    year = "1993"
}

@article{Deur:2018roz,
    author = "Deur, Alexandre and Brodsky, Stanley J. and De T{\'e}ramond, Guy F.",
    title = "{The Spin Structure of the Nucleon}",
    eprint = "1807.05250",
    archivePrefix = "arXiv",
    primaryClass = "hep-ph",
    reportNumber = "JLAB-PHY-18-2760, SLAC-PUB-17279",
    doi = "10.1088/1361-6633/ab0b8f",
    journal = "Rept. Prog. Phys.",
    volume = "82",
    pages = "076201",
    year = "2019"
}

@article{Alexandrou:2026rca,
    author = "Alexandrou, Constantia and Bacchio, Simone and Evangelista, Antonio and Frezzotti, Roberto and Margari, Francesca and Sanfilippo, Francesco and Schneider, Christian",
    title = "{Performance of Low Mode Averaging on Twisted-Mass Fermion Ensembles at the physical pion mass point}",
    eprint = "2606.27072",
    archivePrefix = "arXiv",
    primaryClass = "hep-lat",
    month = "6",
    year = "2026",
    journal = ""
}

\end{document}